\documentclass[journal,twoside]{IEEEtran}
\usepackage[utf8]{inputenc}
\usepackage{amsmath,amssymb,graphicx}
\usepackage{bbm}
\usepackage{xcolor}
\usepackage{cite}
\usepackage[normalem]{ulem} 
\usepackage{graphicx}
\usepackage{subcaption}
\usepackage{cleveref}
\usepackage{rotating}
\usepackage{url}
\usepackage{siunitx}

\newcommand{\ltwo}{\ensuremath{\ell^2}}

\def\Ydet{{Y_{\rm det}}}
\def\Yhigh{{Y_{\rm high}}}
\def\Ylow{{Y_{\rm low}}}

\newcommand{\abs}[1]{\lvert #1 \rvert}

\DeclareFontFamily{U}{mathx}{\hyphenchar\font45}
\DeclareFontShape{U}{mathx}{m}{n}{
      <5> <6> <7> <8> <9> <10>
      <10.95> <12> <14.4> <17.28> <20.74> <24.88>
      mathx10
      }{}
\DeclareSymbolFont{mathx}{U}{mathx}{m}{n}
\DeclareFontSubstitution{U}{mathx}{m}{n}
\DeclareMathAccent{\widecheck}{0}{mathx}{"71}

\newcommand{\Frac}[2]{{{#1}/{#2}}}  

\newcommand{\ds}{\displaystyle}

\newcommand{\beq}{\begin{equation}}
\newcommand{\eeq}{\end{equation}}
\newcommand{\beqan}{\begin{eqnarray*}}
\newcommand{\eeqan}{\end{eqnarray*}}

\newcommand{\eqlabel}[1]{ \stackrel{(#1)}{=} }

\newcommand{\openCase}  {\left\{ \begin{array}{@{\,}ll}}
\newcommand{\openCasell}{\left\{ \begin{array}{@{\,}ll}}
\newcommand{\openCasecl}{\left\{ \begin{array}{@{\,}cl}}
\newcommand{\openCaserl}{\left\{ \begin{array}{@{\,}rl}}
\newcommand{\openCaseTablell}{\left\{ \begin{array}{@{}ll}}
\newcommand{\openCaseTablecl}{\left\{ \begin{array}{@{}cl}}
\newcommand{\openCaseTablerl}{\left\{ \begin{array}{@{}rl}}
\newcommand{\closeCase} {\end{array} \right.}

\def\half{{\textstyle\frac{1}{2}}}

\def\mtilde{\widetilde{m}}
\def\Mtilde{\widetilde{M}}

\def\Ttilde{\widetilde{T}}

\def\xtilde{\widetilde{x}}
\def\Xtilde{\widetilde{X}}

\DeclareMathOperator*{\argmax}{arg\,max}

\newcommand{\iP}[1]{\mathrm{P}({#1})} 
\def\smid{\,|\,}  
\def\sMid{\,;\,}  
\newcommand{\E}[1]{\mathbb{E}\!\left[\,{#1}\,\right]}   
\newcommand{\EP}[1]{\mathrm{E}_P\!\left[\,{#1}\,\right]}   
\newcommand{\iE}[1]{\mathrm{E}[\,{#1}\,]}   
\newcommand{\iEP}[1]{\mathrm{E}_P[\,{#1}\,]}   
\newcommand{\var}[1]{\mathrm{var}\!\left({#1}\right)}   
\newcommand{\normal}[2]{{\mathcal{N}(#1,#2)}}  
\newcommand{\Poisson}[1]{{\mathrm{Poisson}(#1)}}  
\newcommand{\ZIPoisson}[2]{{\mathrm{ZIPoisson}(#1,#2)}}  
\newcommand{\Neyman}[2]{{\mathrm{Neyman}(#1,#2)}}  
\newcommand{\binomial}[2]{{\mathrm{binomial}(#1,#2)}}  
\newcommand{\Beta}[2]{{\mathrm{Beta}(#1,#2)}}  

\def\Utilde{{\widetilde{U}}}

\def\MID{\, \| \,}
\newcommand\DKL[2]{D_{\rm KL}( {#1} \MID {#2} )}

\def\etaML{\widehat{\eta}_{\rm ML}}

\def\PMD{\mathrm{P}_{\rm MD}}
\def\etaQM{\widehat{\eta}_{\rm QM}}
\def\etaLQM{\widehat{\eta}_{\rm LQM}}
\def\etaML{\widehat{\eta}_{\rm ML}}
\def\etaMLI{\widehat{\eta}_{\rm MLI}}
\def\etaCE{\widehat{\eta}_{\rm CE}}
\def\etaCI{\widehat{\eta}_{\rm CI}}
\def\etaIC{\widehat{\eta}_{\rm IC}}
\def\etaICO{\widehat{\eta}_{\rm ICO}}
\def\etaMtilde{\widehat{\eta}_{\Mtilde}}
\def\etaMax{\eta_{\rm max}}

\def\etaconv{\widehat{\eta}_{\rm conv}}
\def\Mtildecorr{\widetilde{M}_{\rm corrected}}

\def\us{\si{\micro\second}}
\def\ns{\si{\nano\second}}

\def\pA{\si{\pico\ampere}}

\def\keV{\si{\kilo\electronvolt}}

\def\V{\si{\volt}}

\newcommand{\etaoracle}{\widehat{\eta}_{\rm oracle}}

\title{Continuous-Time Modeling and Analysis of \\ Particle Beam Metrology}
\author{Akshay Agarwal, Minxu Peng,
and
Vivek K Goyal
\thanks{The authors are with the Department of Electrical and Computer Engineering,
Boston University, Boston, MA 02215 USA (e-mail: akshayag@bu.edu; mxpeng@bu.edu; v.goyal@ieee.org).}%
\thanks{This work was supported in part by the US National Science Foundation under Grant No.\ 1815896 and Grant No.\ 2039762.}
}

\begin{document}

\maketitle

\begin{abstract}
Particle beam microscopy (PBM) performs nanoscale imaging by pixelwise capture of scalar values representing noisy measurements of the response from secondary electrons (SEs) integrated over a dwell time.
Extended to metrology, goals include estimating SE yield at each pixel and detecting differences in SE yield across pixels;
obstacles include shot noise in the particle source as well as lack of knowledge of and variability in the instrument response to single SEs.
A recently introduced time-resolved measurement paradigm promises mitigation of source shot noise,
but its analysis and development have been largely limited to estimation problems under an idealization in which SE bursts are directly and perfectly counted.
Here,
analyses are extended to error exponents in feature detection problems
and to degraded measurements that are representative of actual instrument behavior for estimation problems.
For estimation from idealized SE counts, insights on existing estimators and a superior estimator are also provided.
For estimation in a realistic PBM imaging scenario, extensions to the idealized model are introduced, methods for model parameter extraction are discussed, and
large improvements from time-resolved data are presented.

\end{abstract}


\section{Introduction}

Particle beam microscopy (PBM) techniques such as scanning electron microscopy (SEM)~\cite{Oatley:82,McMullan1995} and helium ion microscopy (HIM)~\cite{erwin1969field,WardNE:06} are widely used to image and characterize samples at the nanoscale.
Images are formed one pixel at a time by raster scanning a focused beam of high-energy charged particles (electrons in SEM and helium ions in HIM) and detecting secondary electrons (SEs) emitted from the sample.
The pixel value is a noisy measurement of the intensity of an SE signal integrated over some \emph{dwell time}.
The scale is often arbitrary;
the micrograph is then an image showing spatial variations without representing a quantified physical property.
Calibration of the \emph{beam current}
(expressed as the mean number of incident particles per unit time)
and the mean instrument response per SE enables the more ambitious goal  of \emph{metrology},
with pixel values representing estimates of \emph{SE yield} per incident particle.

Randomness of the incidence of primary particles---\emph{source shot noise}---is a key characteristic of PBM that contributes to its noise and hence to the amount of averaging that is needed to produce high-quality images.
The achievable image quality in PBM is often limited by the imaging dose
(\textit{i.e.}, number of incident particles).
This limitation is particularly important for radiation-sensitive materials, such as proteins and biomolecules, which are increasingly being imaged by various PBM techniques \cite{Joens:13, Merolli2022}.
Although there has been previous work on the trade-off between dose and image quality~\cite{CastaldoHKVM:09,CastaldoHK:11,Dahmen2016},
as well as attempts to improve PBM image quality through the use of denoising and deconvolution techniques~\cite{BarlowPSC:16,PengCDIKKG:21,Vanderlinde2007,Pang2021,Dahmen2018},
there is a lack of fundamental work around the relationship between information about the sample SE yield and the imaging dose used, as well as a lack of statistically-motivated SE yield estimation techniques based on the signals collected on PBMs.

In this paper, we explore the fundamental limits to particle beam metrology and describe a novel imaging scheme where the integration over a dwell time is replaced with time-resolved (TR) measurement using analog outcoupling of the SE signal.
For both detection and estimation of SE yield,
we find that
TR measurement can lead to
large improvements.
Starting with an idealized model in which SEs are counted perfectly,
we show that the improvements are characterized by
differences in Kullback--Leibler divergence
and Fisher information
between Gaussian and zero-inflated Poisson distributions.
We extend our modeling and analysis to include three causes for
inaccuracy in SE counts:
saturation in SE counting,
additive noise from the detection signal chain,
and
overlap of responses from temporally adjacent incident particles.

The concept of benefiting from time resolution in PBM was introduced in~\cite{PengMBBG:20}.
Continuous-time modeling of PBM was introduced in~\cite{PengMBG:21},
along with theoretical analyses and Monte Carlo simulations of several estimators for SE yield.
Robustness to unknown beam current was shown in~\cite{Watkins2021a,Watkins2021b},
and joint estimation of beam current and SE yield was studied in~\cite{Seidel2022TCI,Seidel2022_MM}.
A recent manuscript develops denoising  procedures to apply with time-resolved data based on plug-and-play methods~\cite{PengKSYG:23arXiv}.
All these previous works concentrate on a model in which SEs are counted perfectly.
Thus, they can give the impression that the benefits of TR measurement are contingent on this idealization.
Here, by including various degradations to SE observation,
we highlight that the benefits from TR measurement 
persist with non-ideal SE detection.

\subsection{Contributions}
The main contributions of this paper include:
\begin{itemize}
    \item \emph{Detection of SE yield.}
    We provide the first results on hypothesis testing between two SE yield values
    from SE-count measurements.
    The unbounded improvements in error exponents due to TR measurement are analyzed using
    Kullback--Leibler divergence.
    \item \emph{Improved estimation of SE yield.}
    We introduce a new estimator for SE yield from SE-count measurements that improves upon the estimators analyzed and simulated in~\cite{PengMBG:21}.
    We also provide new insights on some estimators and bounds in~\cite{PengMBG:21}.
    \item \emph{Estimation from binary SE measurements.}
    We show that SE yield can be estimated from measurements that saturated at 1 SE,
    and we characterize the performance limits under this limited form of measurement.
    \item \emph{Estimation from degraded SE measurements.}
    We analyze the increases of estimation error lower bounds that result from additive noise in SE measurements.
    We also provide a procedure to fit model parameters to experimental data.
    \item \emph{Impact of nonzero pulse width.}
    We introduce a compensation for the possible undercounting of detection events
    due to overlapping of pulses.
\end{itemize}

\subsection{Outline}
\Cref{sec:meas-models} introduces an abstract model for PBM and gathers several preliminary computations
pertaining to the Neyman Type A observations generated with perfect counting of SEs.
\Cref{sec:detection} is dedicated to feature detection abstracted as a binary hypothesis test.
We compute error exponents for conventional and time-resolved measurements,
and we find that the increase of error exponents
(decrease of error probabilities)
with TR measurements is by a large
(potentially unbounded)
factor.
\Cref{sec:estimation} turns to estimation problems, still assuming perfect counting of SEs.
We provide insights into existing estimators and introduce a new estimator that is
based on the conditional expectation of an oracle estimator.
\Cref{sec:saturated-SEs} introduces an estimator for SE yield that does not require SE counts;
instead, it uses only the number of detection events, as one would obtain if the SE detector
saturates at a single electron.
\Cref{sec:ppg} develops a richer model for noisy SE detection and several estimators
to apply in this setting.
We show how the Fisher information of the measurements decays with increasing noise.
We develop methods to fit model parameters and show results with data collected from a real instrument.
Estimation simulations show large improvements from TR measurement.
\Cref{sec:conclusion} concludes.

\section{Abstract Model with Ideal SE Counting}
\label{sec:meas-models}

After briefly describing the operation of a typical instrument in \Cref{sec:typical-instrument},
we review an abstraction that assumes ideal counting of SEs~\cite{PengMBG:21} in \Cref{subsec:stochastic_abstraction}.
This idealization is used for detection problems in \Cref{sec:detection} and for estimation problems in \Cref{sec:estimation}.

The Neyman Type A distribution of an idealized conventional measurement
of SEs, 
developed in \Cref{sec:measurement-distribution},
can be used for various numerical evaluations but is not conducive to closed-form analytical results.
We thus introduce high- and low-dose approximations and
a proxy inspired by the concept of a deterministic incident particle beam.
The approximations and their asymptotes offer insightful intuitions in understanding the behavior of the distribution of the idealized conventional measurement. 
To later aid in contrasting with TR measurement, we also provide
computations of Fisher information (\Cref{sec:background-FI})
and Kullback--Leibler divergence (\Cref{sec:background-KLD})
for the conventional measurement.

While the incident particles may be electrons or ions, for simplicity we refer to them as ions.

\subsection{Operation of a Typical Instrument}
\label{sec:typical-instrument}
In a typical particle-beam imaging setup, shown schematically in \Cref{fig:marked-process-illustration}(a), the SEs emitted from each pixel of the sample are detected by an Everhart-Thornley (ET) detector~\cite{Everhart1960}, which consists of a scintillator followed by a photomultiplier tube (PMT).
The detection of SEs from a single incident ion typically occurs within a few femtoseconds~\cite{LiMD:19},
whereas the mean interarrival time for ions is on the order of 100 $\ns$.
After emission from the sample pixel, the SEs are typically accelerated to 10 $\keV$ and made incident on a scintillator.
The scintillator generates a random number of photons, with the mean proportional to the number of incident SEs.
These photons are then directed towards the PMT though a light pipe, where they generate a voltage pulse with a mean height proportional to their number. Therefore, the final output signal from the ET detector consists of a series of voltage pulses, as depicted by the experimental data shown in \Cref{fig:marked-process-illustration}(e). 

Although the ideal SE image would be a pixel-wise map of the sample SE yield, conventional PBM does not attempt to create such an image due to two factors.
First, the gains and loss factors involved in the SE detection chain are usually not available to the microscopist or the imaging software.
Second, there can be a large variance in the voltage signal generated by the one SE\@.
Due to the lack of knowledge of the mean instrument response per SE, the count of SEs per pixel is conventionally not evaluated during imaging, preventing estimation of the SE yield.
Instead, the voltage signal is sampled at a fixed period (typically 100~ns) and summed for each pixel dwell time to generate a scalar 8-bit pixel brightness.

\subsection{Stochastic Process Abstraction}
\label{subsec:stochastic_abstraction}
\begin{figure}
    \includegraphics[width=\linewidth]{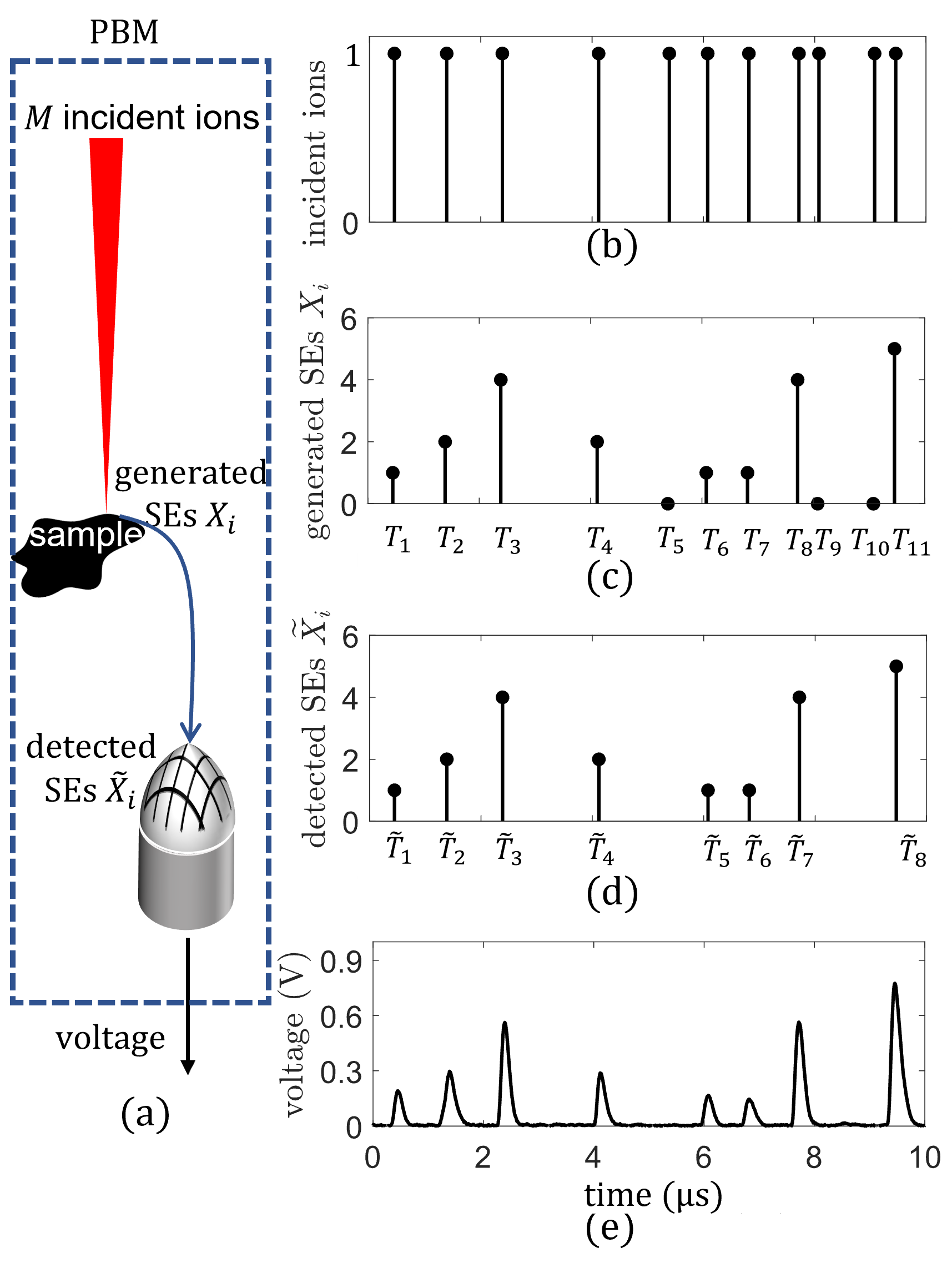}
    \caption{A possible realization of the random processes involved in
    a generative model of SE imaging in PBM\@.
    (a)~Schematic for SE imaging in PBM\@.
    (b)~Generation of $M$ incident ions.
    (c)~The underlying marked Poisson process
    $\{(T_1,X_1),\, (T_2,X_2),\, \ldots \}$
    with ions incident at times $T_1,\,T_2,\,\ldots$
    generating detected SE counts $X_1,\,X_2,\,\ldots$.
    (d)~The marked Poisson process
    $\{(\Ttilde_1,\Xtilde_1),\, (\Ttilde_2,\Xtilde_2),\, \ldots \}$
    produced by discarding the ions for which no SEs are detected.
    (e)~SE detector voltage response.
    Panel is a real snapshot of voltage output from an HIM.}
    \label{fig:marked-process-illustration}
\end{figure}

Our measurement model and estimation techniques are separable across the pixels,
so we omit any pixel indexing.
Denote the pixel dwell time by $t$. 
For each pixel, the incident ion arrivals are modeled as a Poisson process with known rate $\Lambda$ per unit time,
as illustrated in \Cref{fig:marked-process-illustration}(b) for $t=10~\us$.
Incident ion $i$ interacts with the sample, causing $X_i$ number of SEs to be detected,
as illustrated in \Cref{fig:marked-process-illustration}(c).
Each $X_i$ can be described as a Poisson random variable with mean $\eta$~\cite{Joy2008}. 
This $\eta$ is called the \emph{SE yield} and is the parameter we wish to measure for the pixel.
Note that detection efficiency~\cite{Joy1996,Joy2006} is incorporated within the definition of $\eta$.

The model can be described as a marked Poisson process
$\{(T_1,X_1),\, (T_2,X_2),\, \ldots \}$,
where
$(T_1,\, T_2,\, \ldots)$ is the arrival time sequence of the ions.
The number of incident ions $M$ is the largest $i$
such that $T_i \leq t$
(with $M=0$ when $T_1 > t$). 
This $M$ is a Poisson random variable with mean $\lambda = \Lambda t$, which we call the dose.%
\footnote{More commonly, dose is the mean number of incident particles per unit area; here, we are not considering absolute spatial scale.}
Ions are observed only indirectly through the detection of SEs.
There is no observed event when $X_i = 0$.
Hence what is observable is a marked thinned Poisson process
$\{(\Ttilde_1,\Xtilde_1),\, (\Ttilde_2,\Xtilde_2),\, \ldots \}$,
where
$\Ttilde_i$ is the arrival time of the $i$th ion that produces a \emph{positive} number of detected SEs
and $\Xtilde_i$ is the corresponding number of detected SEs,
as illustrated in \Cref{fig:marked-process-illustration}(d).
Define $\Mtilde$ to be the largest $i$ such that $\Ttilde_i \leq t$
(with $\Mtilde=0$ when $\Ttilde_1 > t$).

\subsection{Time-Resolved Measurement Model}
\label{sec:tr-model}
Observation of 
\begin{align}
    \left\{ \Mtilde, \Ttilde, \Xtilde \right\} 
    &= \left\{ \Mtilde,
       (\Ttilde_1,\Ttilde_2,...,\Ttilde_{\Mtilde}),
       (\Xtilde_1,\Xtilde_2,...,\Xtilde_{\Mtilde}) \right\}
\label{eq:CTTR-observation}
\end{align}
was introduced in~\cite{PengMBG:21} as \emph{continuous-time time-resolved} measurement,
contrasting with a discrete-time model introduced earlier in~\cite{PengMBBG:20}.
Here we will consider only the continuous-time setting,
which facilitates simpler and more easily interpretable results.

Since the thinning is independent of the ion incidence process
and $\iP{X_i = 0} = e^{-\eta}$,
\begin{equation}
   \label{eq:Mtilde-distribution}
   \Mtilde \sim \Poisson{\lambda(1-e^{-\eta})}.
\end{equation}
Each $\Xtilde_i$ has the zero-truncated Poisson distribution with parameter $\eta$:
\begin{equation}
   \label{eq:Xtilde-distribution}
   \mathrm{P}_{\Xtilde_i}(j ; \eta) = \frac{e^{-\eta}}{1-e^{-\eta}} \cdot \frac{\eta^j}{j!},
   \qquad 
   j = 1,\, 2,\, \ldots.
\end{equation}
The mean of this distribution is
\begin{equation}
    \label{eq:Xtilde-mean}
    \E{ \Xtilde_i } = \frac{\eta}{1 - e^{-\eta}}.
\end{equation}
Given $\Mtilde$, 
$(\Xtilde_1,\,\ldots,\,\Xtilde_{\Mtilde})$
are independent and identically distributed.
Conditioned on $\Mtilde = \mtilde > 0$,
the normalized time $\Ttilde_i/t$ has the $\Beta{i}{\mtilde+1-i}$ distribution
(with no dependence on $\eta$).

\subsection{Conventional Measurement Distribution}
\label{sec:measurement-distribution}
As discussed in \Cref{sec:typical-instrument},
a typical instrument generates a single scalar value for each raster scan location,
and this value has many sources of noise.
An idealized scalar measurement is for the instrument to give the cumulative SE counts within dwell time $t$
(\textit{i.e.}, the measurement results in a \emph{scalar} value for every pixel, as opposed to vector-valued TR measurement):
\begin{equation}
\label{eq:Y-def}
    Y = \sum_{i=i}^M X_i.
\end{equation}
This $Y$ is a Neyman Type A random variable with parameters $\lambda$ and $\eta$,
which we will denote $\Neyman{\lambda}{\eta}$.
Its probability mass function (PMF) is
\begin{equation}
\label{equ:neyman}
    \mathrm{P}_{Y}(y \sMid \eta, \lambda) = \frac{e^{-\lambda} \eta^y}{y!} \sum_{m = 0}^{\infty} \frac{(\lambda e^{-\eta}) ^m m^y}{m!},
    \quad
    y = 0,\,1,\,\ldots,
\end{equation}
its mean is
\begin{equation}
\label{eq:neyman_mean}
    \E{Y} = \lambda\eta,
\end{equation}
and its variance is
\begin{equation}
\label{eqn:neyman_var}
    \var{Y} = \lambda\eta(\eta + 1).
\end{equation}
Like for a Poisson distribution,
the variance increases with the mean;
unlike a Poisson distribution, the variance exceeds the mean,
and this is increasingly true as $\eta$ increases.
This excess variance is consistent with experimental observations, and compound Poisson distributions have been previously used to model the distribution of SEs in PBM~\cite{Uchikawa1992, Frank2005, Novak2009a, TimischlDN:12}.

The series appearing within the PMF \eqref{equ:neyman}
makes the Neyman Type A distribution difficult to work with both
analytically and computationally.
While we will sometimes use \eqref{equ:neyman} directly,
it simplifies some computations and makes certain comparisons more
intuitive to use approximations that hold for high or low $\lambda$.
A purely hypothetical situation of a deterministic incident beam
also provides valuable context.

\subsubsection{Deterministic beam (Poisson approximation)}
If $\lambda$ is a positive integer,
we may imagine a situation in which exactly $\lambda$ ions are incident.
Since the sum of independent Poisson random variables is a Poisson random variable,
we obtain a simple model of
\begin{equation}
  \label{eq:PoissonApproximation}
    \Ydet \sim \Poisson{\lambda \eta}.
\end{equation}
Notice
by comparison to \eqref{eqn:neyman_var}
that $Y$ has higher variance than $\Ydet$ by a factor of $\eta + 1$.
This $\eta + 1$ factor is attributed to the randomness of the incident ion counts,
\textit{i.e.}, source shot noise.

\subsubsection{High $\lambda$ (Gaussian approximation)}
As $\lambda \rightarrow \infty$,
a Gaussian approximation with matching moments in \eqref{eq:neyman_mean} and \eqref{eqn:neyman_var}
holds
in the sense of pointwise convergence of moment generating functions~\cite[$\S$2a]{MartinK:62}:
\begin{equation}
  \label{eq:GaussianApproximation_PDF}
    \Yhigh \sim \normal{\lambda \eta}{\lambda\eta(\eta+1)}.
\end{equation}
One may use \eqref{eq:GaussianApproximation_PDF} to form an approximate PMF
by integrating over intervals $\{[y-\half,\,y+\half]\}_{y=0}^\infty$.
For $\lambda > 10$ and $\eta > 1$,
the squared $\ltwo$ error of this approximation is less than 0.002~\cite[Fig.~1]{MartinK:62}.
Dose exceeding 10 ions per pixel is typical for useful micrograph quality.

\subsubsection{Low $\lambda$ (zero-inflated Poisson approximation)}
As $\lambda \rightarrow 0$,
the PMF \eqref{equ:neyman} converges pointwise to a Poisson distribution
with extra mass at zero~\cite[$\S$2b]{MartinK:62}:
\begin{equation}
    \label{eq:NeymanA-low-lambda}
    \mathrm{P}_{\Ylow}(y \sMid \eta, \lambda) 
         = \openCaserl
     e^{-\lambda} + (1-e^{-\lambda})e^{-\eta}, & y = 0; \\
          (1-e^{-\lambda})e^{-\eta} \eta^y/y!, & y = 1,\,2,\,\ldots.
          \closeCase
\end{equation}
We will denote this $\ZIPoisson{\lambda}{\eta}$.
For $\lambda < 0.3$ and $\eta < 10$,
the squared $\ltwo$ error of this approximation is less than 0.001~\cite[Fig.~1]{MartinK:62}.
Our use of this approximation is to understand continuous-time behavior,
where $\lambda$ is effectively infinitesimal.

\subsection{Fisher Information}
\label{sec:background-FI}
Fisher information (FI) is a basic tool for lower bounding the mean-squared errors
of estimators.
Here we gather computations of FI that will be used to
contextualize the FI of TR measurements.

\subsubsection{Deterministic beam (Poisson approximation)}
The Fisher information about mean $\nu$ in a $\Poisson{\nu}$ observation is
\begin{equation*}
\mathcal{I}(\nu) = \frac{1}{\nu}.
\end{equation*}
Thus, we have
$\mathcal{I}_\Ydet(\lambda \eta) = \Frac{1}{(\lambda \eta)}$.
With $\lambda$ known, this translates by simple rescaling to

\begin{equation}
\label{eq:FI_Poisson}
\frac{1}{\lambda}{\mathcal{I}_\Ydet(\eta; \lambda)} = \frac{1}{\eta}
\end{equation}
in a normalized form we will use below.

\subsubsection{High $\lambda$ (Gaussian approximation)}
The Fisher information about mean $\mu$ in a Gaussian 
$\normal{\mu}{\sigma^2}$ observation is
\begin{equation*}
\mathcal{I}(\mu)
  = \frac{1}{\sigma^2}.
\end{equation*}
Using the Gaussian approximation \eqref{eq:GaussianApproximation_PDF} to the $\Neyman{\lambda}{\eta}$ distribution
suggests heuristically that the Fisher information about $\lambda \eta$ in $\Yhigh$ is
\begin{equation}
\mathcal{I}_{\Yhigh}(\lambda \eta) = \frac{1}{\lambda\eta(\eta+1)}.
\end{equation}
With $\lambda$ known, this translates by simple rescaling to
\begin{equation}
\frac{1}{\lambda} \mathcal{I}_{\Yhigh}(\eta; \lambda) = \frac{1}{\eta(\eta+1)}
                             = \left(\frac{1}{\eta}-\frac{1}{\eta+1}\right).
\end{equation}
Indeed a detailed argument for
\begin{equation}
\label{eq:FI_high_lambda}
    \lim_{\lambda \to \infty} \frac{1}{\lambda} {\mathcal{I}_Y(\eta;\lambda)} = \frac{1}{\eta} - \frac{1}{\eta + 1}
\end{equation}
is given in~\cite[App.~B]{PengMBG:21}.
Comparing to \eqref{eq:FI_Poisson}, the Fisher information is reduced by a factor of $\eta + 1$.

\subsubsection{Low $\lambda$ (zero-inflated Poisson approximation)}
From the PMF \eqref{eq:NeymanA-low-lambda},
\begin{align}
    &\log \mathrm{P}_{\Ylow}(y \sMid \eta, \lambda) \nonumber \\ 
         &\,\,= \openCaserl
     \log(e^{-\lambda} + (1-e^{-\lambda})e^{-\eta}), & y = 0; \\
          \log(1-e^{-\lambda}) -\eta + y \log \eta - \log(y!), & y = 1,\,2,\,\ldots.
          \closeCase
    \nonumber
\end{align}
Differentiating gives
\begin{align}
    &\frac{\partial \log \mathrm{P}_{\Ylow}(y \sMid \eta, \lambda)}{\partial \eta} \nonumber \\ 
         &\,\,= \openCaserl
     \ds \frac{(1-e^{-\lambda})e^{-\eta}}
              {e^{-\lambda} + (1-e^{-\lambda})e^{-\eta}}, & y = 0; \\
                                          -1 + y / \eta , & y = 1,\,2,\,\ldots.
          \closeCase
    \label{eq:NeymanA-low-lambda-log-PMF-derivative}
\end{align}
Now computing the expected value of the square of this quantity under the
PMF \eqref{eq:NeymanA-low-lambda} gives
\begin{equation}
    \label{eq:ZeroInflatedFisherInformation}
   \mathcal{I}_{\Ylow}(\eta;\lambda) = 
     \frac{(1-e^{-\lambda})^2e^{-2\eta}}{e^{-\lambda} + (1-e^{-\lambda})e^{-\eta}} + (1-e^{-\lambda})\left(\frac{1}{\eta} - e^{-\eta}\right) .
\end{equation}

In the limit of low $\lambda$, the first term approaches zero and the first factor of the second term approaches $\lambda$, so
\begin{equation}
    \label{eq:FI_low_lambda}
   \lim_{\lambda \to 0} \frac{1}{\lambda} {\mathcal{I}_{\Ylow}(\eta;\lambda)} = \frac{1}{\eta} - e^{-\eta} .
\end{equation}
This matches a more tedious derivation of
\begin{equation}
    \label{eq:FI_low_lambda_exact}
   \lim_{\lambda \to 0} \frac{1}{\lambda} {\mathcal{I}_{Y}(\eta;\lambda)} = \frac{1}{\eta} - e^{-\eta} 
\end{equation}
in~\cite[App.~B]{PengMBG:21}.

The low-$\lambda$ limit of Fisher information in \eqref{eq:FI_low_lambda} exceeds the high-$\lambda$ limit in \eqref{eq:FI_high_lambda} by a factor of
$(\eta+1)(1-\eta e^{-\eta})$.
This factor varies from 1 when $\eta =0$ to $\approx \eta + 1$ when $\eta$ is high.
This gain in Fisher information can be attributed to increasing certainty in the number of incident ions at low $\lambda$
and consequent reduction in source shot noise~\cite{PengMBBG:20}. 

\subsection{Kullback--Leibler Divergence}
\label{sec:background-KLD}
Kullback-Leibler divergence (KLD) is a basic tool for quantifying distances between distributions
and in particular determining error exponents for hypothesis testing.
Here we gather computations of KLD
applicable to distinguishing $\Neyman{\lambda}{\eta_0}$ and $\Neyman{\lambda}{\eta_1}$ distributions.
This will be used to contextualize the KLD of TR measurements.

For distributions $p$ and $q$ on the same alphabet,
the Kullback--Leibler divergence is
\begin{equation}
    \label{eq:KL}
\DKL{p}{q} = \EP{\log(p(Y)/q(Y))},
\end{equation}
which is a shorthand for the expected value
of the random variable $\log(p(Y)/q(Y))$
when $Y$ has the $p$ distribution.

\subsubsection{Deterministic beam (Poisson approximation)}
For generic Poisson distributions, the KLD is given by
\begin{equation}
    \DKL{ \Poisson{\nu_0} }
        { \Poisson{\nu_1} }
    = \nu_1 - \nu_0 + \nu_0 \log \frac{\nu_0}{\nu_1}.
    \label{eq:KL-Poisson}
\end{equation}
Thus, we have  
\begin{align}
    &\frac{1}{\lambda} \DKL{ \Poisson{\lambda \eta_0} }
                           { \Poisson{\lambda \eta_1} } \nonumber \\
        &\,\,= \frac{1}{\lambda} \left[ \lambda\eta_1 - \lambda\eta_0 + \lambda\eta_0 \log \frac{\lambda\eta_0}{\lambda\eta_1} \right] \nonumber \\
        &\,\,= \eta_1 - \eta_0 + \eta_0 \log \frac{\eta_0}{\eta_1}.
    \label{eq:KL-deterministic}
\end{align}

\subsubsection{High $\lambda$ (Gaussian approximation)}
For generic univariate Gaussian distributions, the KLD is given by
\begin{align}
    &\DKL{ \normal{\mu_0}{\sigma_0^2} }
         { \normal{\mu_1}{\sigma_1^2} } \nonumber \\
        &\,\,= \frac{1}{2} \log\frac{\sigma_1^2}{\sigma_0^2} + \frac{\sigma_0^2 + (\mu_0-\mu_1)^2}{2\sigma_1^2}-\frac{1}{2}.
    \label{eq:KL-Gaussian}
\end{align}
Thus, we have
\begin{align}
     &\DKL{ \normal{\lambda \eta_0}{\lambda \eta_0(\eta_0+1)} }
          { \normal{\lambda \eta_1}{\lambda \eta_1(\eta_1+1)} } \nonumber \\
        &\,\,= \frac{1}{2} \log\frac{\lambda \eta_1(\eta_1+1)}
                                {\lambda \eta_0(\eta_0+1)}
              + \frac{\lambda \eta_0 (\eta_0+1) + (\lambda \eta_0 - \lambda \eta_1)^2}
                     {2 \lambda \eta_1 (\eta_1 + 1)}
              - \frac{1}{2}
                   \nonumber \\
        &\,\,= \frac{1}{2} \log\frac{\eta_1(\eta_1+1)}
                                {\eta_0(\eta_0+1)}
              + \frac{\eta_0 (\eta_0+1) + \lambda(\eta_0 - \eta_1)^2}
                     {2 \eta_1 (\eta_1 + 1)}
              - \frac{1}{2}.
    \label{eq:GaussianApproximation}
\end{align}
Furthermore,
\begin{align}
    \label{eq:Gaussian-limit}
    \lim_{\lambda \rightarrow \infty}&
   \frac 
   {
              \DKL{ \normal{\lambda \eta_0}{\lambda \eta_0(\eta_0+1)} }
                  { \normal{\lambda \eta_1}{\lambda \eta_1(\eta_1+1)} } 
   }
   {\lambda} \nonumber \\
   &= \frac{ (\eta_0-\eta_1)^2 }
          { 2 \eta_1 (\eta_1 + 1) }.
\end{align}

\subsubsection{Low $\lambda$ (zero-inflated Poisson approximation)}
For zero-inflated Poisson distributions following
\eqref{eq:NeymanA-low-lambda},
we can compute the KLD
$\DKL{ \ZIPoisson{\lambda}{\eta_0} }{ \ZIPoisson{\lambda}{\eta_1} }$
directly.
For $y=0$ we have
\begin{equation}
    \label{eq:log-ratio-0}
   \log \frac{p(0)}{q(0)}
   =
   \log \frac{ e^{-\lambda} + (1-e^{-\lambda})e^{-\eta_0} }
             { e^{-\lambda} + (1-e^{-\lambda})e^{-\eta_1} } ;
\end{equation}
for $y = 1,\,2,\,\ldots$, we have
\begin{equation*}
    \label{eq:ratio-nonzero}
   \frac{p(y)}{q(y)}
   = \frac{ (1-e^{-\lambda}) e^{-\eta_0} \eta_0^y / y! }
          { (1-e^{-\lambda}) e^{-\eta_1} \eta_1^y / y! }
   = e^{-(\eta_0-\eta_1)} (\eta_0/\eta_1)^y,
\end{equation*}
so
\begin{equation}
    \label{eq:log-ratio-nonzero}
   \log \frac{p(y)}{q(y)}
   = \eta_1-\eta_0 + y \log \frac{ \eta_0 }{ \eta_1 }.
\end{equation}
For the KLD, we would like to average \eqref{eq:log-ratio-0} and \eqref{eq:log-ratio-nonzero} under the $p$ distribution:

\begin{subequations}
    \label{eq:LowLambdaApproximation}
\begin{align}
&\DKL{ \ZIPoisson{\lambda}{\eta_0} }{ \ZIPoisson{\lambda}{\eta_1} } \nonumber \\
&\,\,= g(\eta_0) \log \frac{ g(\eta_0) } { g(\eta_1) } \nonumber\\
&\qquad+ (1 - g(\eta_0)) ( \eta_1 - \eta_0 ) + (1 - e^{-\lambda}) \eta_0 \log \frac{\eta_0}{\eta_1},
    \label{eq:LowLambdaApproximation1}
\end{align}
where
\begin{equation}
    \label{eq:g}
    g(s) = e^{-\lambda} + (1-e^{-\lambda})e^{-s}.
\end{equation}
\end{subequations}

Furthermore,
\begin{align}
   &\lim_{\lambda \rightarrow 0}
   \frac 
   {
    \DKL{ \ZIPoisson{\lambda}{\eta_0} }
        { \ZIPoisson{\lambda}{\eta_1} }
   }
   {\lambda} \nonumber \\
   &\,\,= e^{-\eta_0} - e^{-\eta_1}
     + (1-e^{-\eta_0})(\eta_1 - \eta_0)
     + \eta_0 \log \frac{\eta_0}{\eta_1}.
    \label{eq:LowLambdaPoissonLike-limit}
\end{align}

Comparing \eqref{eq:KL-deterministic},
\eqref{eq:Gaussian-limit},
and
\eqref{eq:LowLambdaPoissonLike-limit}
is more subtle than the analogous comparison of FI expressions.
We defer this to the following section in the context of specific numerical examples of error exponents.

\section{Feature Detection with SE Counting}
\label{sec:detection}

A common goal in PBM is to decide on the presence or absence of a feature
that is revealed by deviation of SE yield $\eta$ from the value of surrounding pixels.
For instance, detecting feature positions is a crucial step in improving the accuracy of line-edge roughness measurement~\cite{croon2002line}, which can be helpful in assessing semiconductor manufacturing accuracy.
Here we consider feature detection when SE count data is available as described in \Cref{sec:meas-models}.

\subsection{A Binary Hypothesis Test}
To illustrate the fundamental advantage of TR measurements for feature detection,
we consider a binary hypothesis testing problem
between SE yield values of $\eta_0$ (``no alarm'') and $\eta_1$ (``alarm''),
with dose $\lambda$ known.
With the (idealized) conventional measurement $Y$,
the decision must be made based on whether the observation more plausibly came from the
$\Neyman{\lambda}{\eta_0}$
or
$\Neyman{\lambda}{\eta_1}$
distribution.
Observation of 
$\{ \Mtilde,\, (\Ttilde_1,\Xtilde_1),\, (\Ttilde_2,\Xtilde_2),\, \ldots,\, (\Ttilde_{\Mtilde},\Xtilde_{\Mtilde}) \}$
is at least as informative,
and we wish to characterize how much the decision making accuracy is improved.

In this section, we imagine that a PBM experiment with dose $\lambda$ is repeated many times.
We study the performance through the rate of exponential decay of the missed detection rate
for a sequence of Neyman--Pearson hypothesis tests that minimize the missed detection rate
while satisfying a fixed false alarm rate criterion.
For $n$ repetitions, the probability of missed detection $\PMD(n)$ satisfies
\begin{equation}
    \lim_{n \rightarrow \infty} -\frac{1}{n} \log \PMD(n) = \DKL{p_0}{p_1},
\end{equation}
where $p_0$ and $p_1$ represent the relevant observation distributions~\cite[$\S$8.3.2]{MoulinV:19}.
Thus, we concentrate on KLD computations and comparisons.

\subsection{KLD Between Time-Resolved Measurement Distributions}
\label{sec:KLD-TR}

Let $p$ and $q$ denote the distributions under SE yields $\eta_0$ and $\eta_1$,
with the random variables allowed to be implicit.
In anticipation of using the law of iterated expectation, we make the following simplification:
\begin{align}
   &\EP{\log (p(\Mtilde,\Ttilde,\Xtilde)/q(\Mtilde,\Ttilde,\Xtilde)) \,\big|\, \Mtilde = \mtilde} \nonumber \\
   &\,\,\eqlabel{a} \EP{\log (p(\Mtilde,\Xtilde)/q(\Mtilde,\Xtilde)) \,\big|\, \Mtilde = \mtilde} \nonumber \\
   &\,\,\eqlabel{b} \EP{\log \frac{p(\Mtilde)}{q(\Mtilde)} \,\Big|\, \Mtilde = \mtilde}
                            + \sum_{i=1}^{\mtilde} \EP{ \log \frac{p(\Xtilde_i)}{q(\Xtilde_i)} }
                                    \nonumber \\
   &\,\,\eqlabel{c} \log \frac{p(\mtilde)}{q(\mtilde)}
                            + \mtilde \, \EP{ \log \frac{p(\Xtilde_i)}{q(\Xtilde_i)} },
   \label{eq:conditional-mean-for-kld}
\end{align}
where
(a) follows from the conditional distribution of each $\Ttilde_i$ given $\Mtilde$ being identical under $p$ and $q$;
(b) from the conditional independence of
$\{\Xtilde_1,\, \ldots,\, \Xtilde_{\Mtilde}\}$ given $\Mtilde = \mtilde$;
and
(c) from the $\Xtilde_i$ distributions being identical.
Now by taking the expected value of \eqref{eq:conditional-mean-for-kld}
and using the law of iterated expectation, we obtain
\begin{align}
  &\DKL{p(\Mtilde,\Ttilde,\Xtilde)}{q(\Mtilde,\Ttilde,\Xtilde)} \nonumber \\
  &\,\,= \EP{\log(p(\Mtilde,\Ttilde,\Xtilde)/q(\Mtilde,\Ttilde,\Xtilde))} \nonumber \\
  &\,\,= \DKL{p(\Mtilde)}{q(\Mtilde)} + \iE{\Mtilde} \DKL{p(\Xtilde_i)}{q(\Xtilde_i)}.
   \label{eq:expanded-kld}
\end{align}
Recall the $\Mtilde$ distribution is given in \eqref{eq:Mtilde-distribution}.
Thus, we can apply \eqref{eq:KL-Poisson}
with $\nu_i = \lambda(1-e^{-\eta_i})$
to obtain
\begin{align}
    &\DKL{p(\Mtilde)}{q(\Mtilde)} \nonumber \\
    &\,\,= \lambda(e^{-\eta_0} - e^{-\eta_1}) + \lambda(1 - e^{-\eta_0}) \log \frac{1 - e^{-\eta_0}}{1 - e^{-\eta_1}}.
    \label{eq:KLD-Mtilde}
\end{align}
Recall also the $\Xtilde_i$ distribution is given in \eqref{eq:Xtilde-distribution}.
Thus, we have the log PMF ratio
\begin{align}
    \log \frac{p(x)}{q(x)}
         &= \log \frac{e^{-\eta_0} \eta_0^x / [ (1-e^{-\eta_0}) x!]}
                      {e^{-\eta_1} \eta_1^x / [ (1-e^{-\eta_1}) x!]} \nonumber \\
         &= \eta_1-\eta_0
             + \log \frac{ 1 - e^{-\eta_1} }
                         { 1 - e^{-\eta_0} }
             + x \log\!\left( \frac{\eta_0}{\eta_1} \right).
\end{align}
Taking the expectation under distribution $p$ and using \eqref{eq:Xtilde-mean} gives
\begin{align}
    &\DKL{p(\Xtilde_i)}{q(\Xtilde_i)} \nonumber \\
    &\,\,= \eta_1-\eta_0
             + \log \frac{ 1 - e^{-\eta_1} }
                         { 1 - e^{-\eta_0} }
             + \frac{\eta_0}{1 - e^{-\eta_0}} \log\!\left( \frac{\eta_0}{\eta_1} \right).
    \label{eq:KLD-Xtilde}
\end{align}
Finally, substituting $\iEP{\Mtilde} = \lambda(1-e^{-\eta_0})$,
\eqref{eq:KLD-Mtilde}, and \eqref{eq:KLD-Xtilde} into
\eqref{eq:expanded-kld} gives
\begin{align}
  &\DKL{p(\Mtilde,\Ttilde,\Xtilde)}{q(\Mtilde,\Ttilde,\Xtilde)} \nonumber \\
    &\,\,= \lambda(e^{-\eta_0} - e^{-\eta_1}) + 
                   \lambda(1-e^{-\eta_0})(\eta_1-\eta_0) \nonumber \\
    &\qquad + \lambda \eta_0 \log(\eta_0/\eta_1).
   \label{eq:ct-kld}
\end{align}

Notice that the KLD \eqref{eq:ct-kld} matches the asymptote \eqref{eq:LowLambdaPoissonLike-limit}
of the low-$\lambda$ approximation to $Y$.
This is analogous to a previous known result for FI about $\eta$ normalized by $\lambda$~\cite[Sect.~III-V]{PengMBG:21}:
the normalized FI of a continuous-time TR measurement
matches
the low-$\lambda$ asymptote for a conventional measurement.
The KLD and FI results have an intuitive rationale.
When the dose $\lambda$ is very small, the probability of more than one incident ion is negligible,
and
with zero or one incident ion,
the conventional and TR measurements are identical.
A single low-$\lambda$ conventional measurement is not informative enough for
useful detection or estimation.
The analyses of KLD and FI indicate that TR measurements achieve the best possible
informativeness per incident particle, but uniformly over $\lambda$.

\subsection{Comparisons of Error Exponents}
\label{detect_sec:simulation}

The Neyman Type A PMF \eqref{equ:neyman} is not amenable to meaningful expressions for KLD\@.
If series truncation is handled with care
and overflow and underflow are avoided,
one can numerically evaluate
$\DKL{ \Neyman{\lambda}{\eta_0} } { \Neyman{\lambda}{\eta_1} }$.

\begin{figure}
  \begin{center}
    \begin{tabular}{c}
      \includegraphics[width=0.9\linewidth]{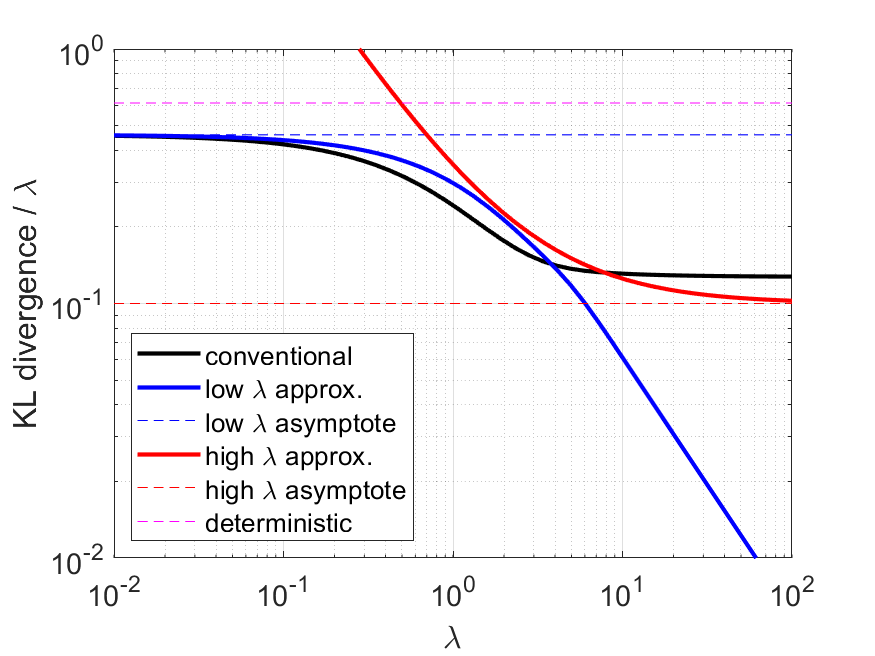} \\
      {\small (a) $\eta_0 = 2$, $\eta_1 = 4$} \\
      \includegraphics[width=0.9\linewidth]{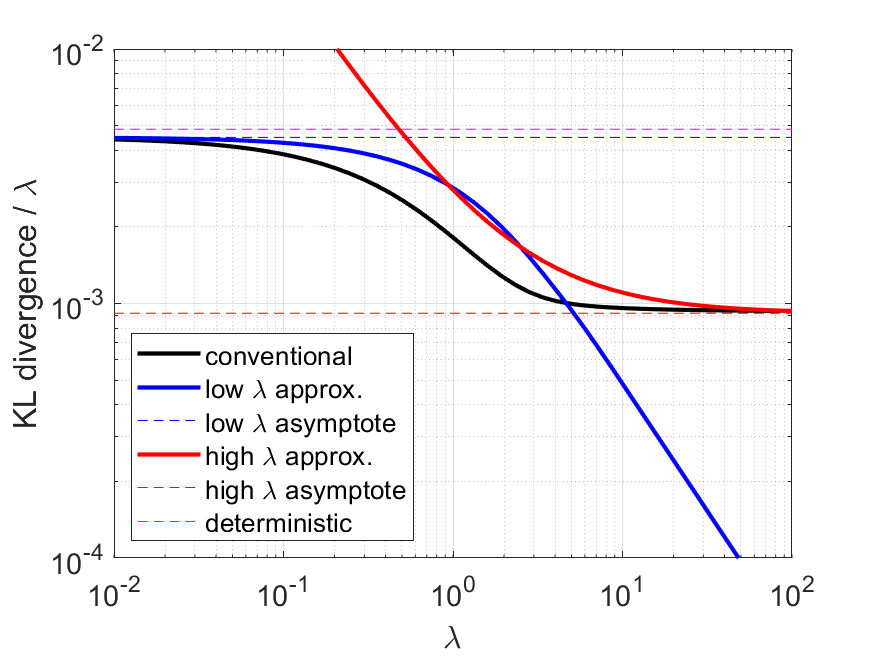} \\
      {\small (b) $\eta_0 = 4$, $\eta_1 = 4.2$} 
    \end{tabular}
  \end{center}
  \caption{
    Computations of normalized KL divergence
    $\DKL{ \Neyman{\lambda}{\eta_0} } { \Neyman{\lambda}{\eta_1} }/\lambda$
    and approximations and asymptotes derived herein.
    Magenta (dashed): deterministic beam \eqref{eq:KL-deterministic}.
    Red:  high-$\lambda$ approximation \eqref{eq:GaussianApproximation}
    and its asymptote \eqref{eq:Gaussian-limit}.
    Blue:  low-$\lambda$ approximation \eqref{eq:LowLambdaApproximation}
    and its asymptote \eqref{eq:LowLambdaPoissonLike-limit}.
    The normalized KL divergence with TR measurement matches the
    low-$\lambda$ asymptote for all values of $\lambda$.
    }
  \label{fig:KL}
\end{figure}

\Cref{fig:KL} provides a few examples.
In black is the normalized KLD
$\DKL{ \Neyman{\lambda}{\eta_0} } { \Neyman{\lambda}{\eta_1} }/\lambda$.
The normalized KLD is also shown for
the deterministic beam (Poisson approximation) \eqref{eq:KL-deterministic};
the high-$\lambda$ (Gaussian approximation) \eqref{eq:GaussianApproximation}
and its asymptote \eqref{eq:Gaussian-limit};
and
the low-$\lambda$ (zero-inflated Poisson approximation) \eqref{eq:LowLambdaApproximation}
and its asymptote \eqref{eq:LowLambdaPoissonLike-limit}.
The normalized KLD with TR measurement \emph{equals} the low-$\lambda$ asymptote.

The most important observation is that the KLD with TR measurement is always greater than
with conventional measurement.
This translates to a larger error exponent
and hence a lower missed detection rate when false alarm rate is held constant.

\begin{figure}
  \begin{center}
    \begin{tabular}{c}
      \includegraphics[width=0.9\linewidth]{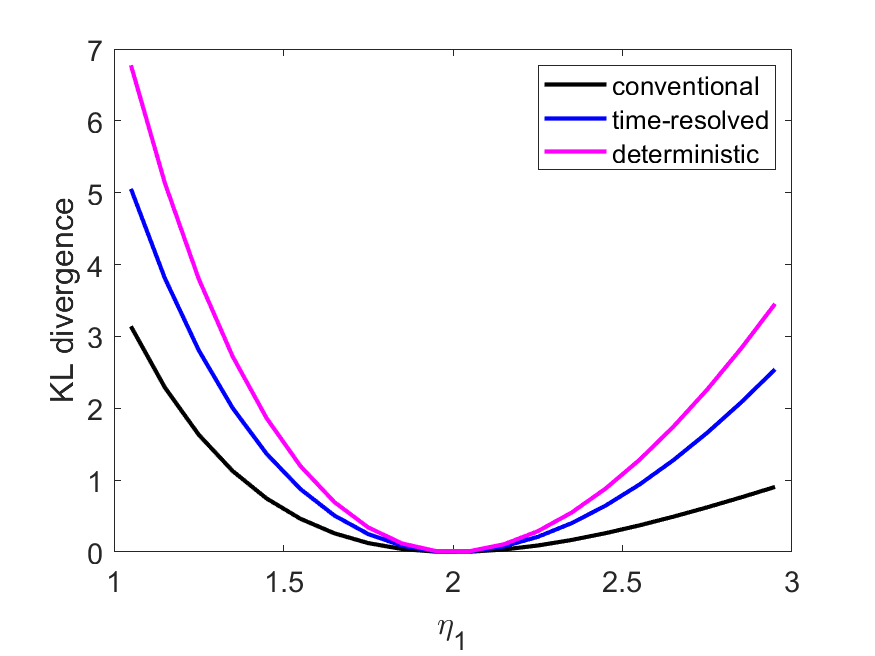} \\
      {\small (a) $\eta_0 = 2$, $\lambda = 20$} \\
      \includegraphics[width=0.9\linewidth]{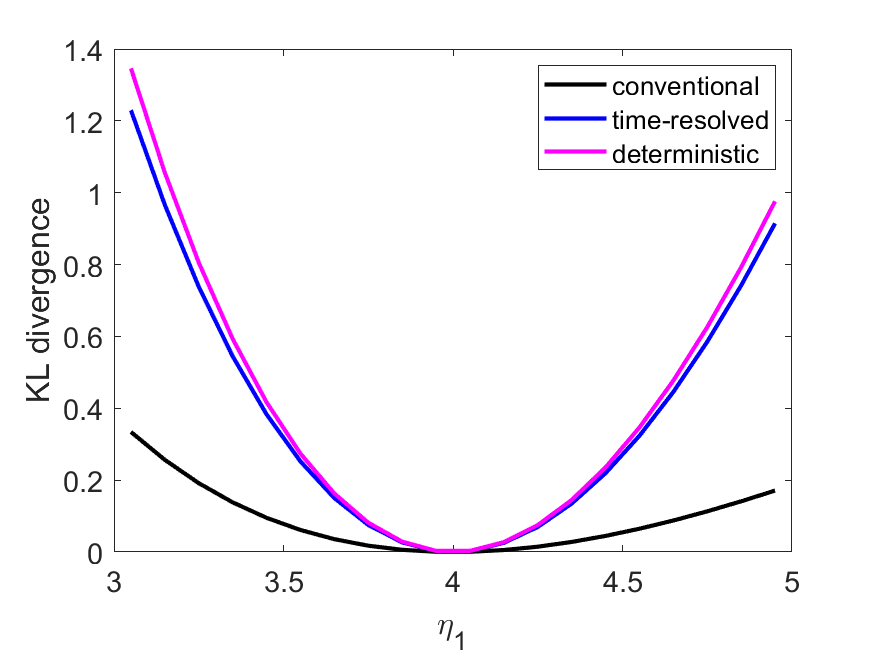} \\
      {\small (b) $\eta_0 = 4$, $\lambda = 20$} 
    \end{tabular}
  \end{center}
  \caption{
    Comparison of error exponents for hypothesis test between $\eta_0$ and $\eta_1$
    where $\eta_0$ is fixed and $\eta_1$ is varied.
    With conventional measurement, the error exponent is
    $\DKL{ \Neyman{\lambda}{\eta_0} } { \Neyman{\lambda}{\eta_1} }$.
    With time-resolved measurement, the error exponent \eqref{eq:ct-kld}
    is significantly larger (superior) and close to the
    error exponent \eqref{eq:KL-deterministic} one would obtained with a
    deterministic source beam.
    }
  \label{fig:KLD_comparison}
\end{figure}

The performance gap can be arbitrarily large.
\Cref{fig:KLD_comparison} compares error exponents over ranges of $\eta_1$ values
for $\eta_0 = 2$ and $\eta_0 = 4$.
The gap is increasing with $\abs{\eta_1-\eta_0}$.
This can also be predicted by comparing \eqref{eq:ct-kld} with \eqref{eq:Gaussian-limit}.

Having established that the error exponent \eqref{eq:ct-kld}
(matching \eqref{eq:LowLambdaPoissonLike-limit})
is achievable,
it is interesting to make additional comparisons.
By comparing \eqref{eq:KL-deterministic} and \eqref{eq:ct-kld},
we see that TR measurement approaches the performance with a deterministic beam
when $\eta_0$ and $\eta_1$ both grow without bound.
In \Cref{fig:KL,fig:KLD_comparison}, the gap is smaller when $\eta_0$ and $\eta_1$ are large.
An interpretation is that when the SE yield is large,
CT measurement allows almost perfect knowledge of the number of incident ions $M$;
evidently, the fact that $M$ is nevertheless random has no impact on the decision
between $\eta_0$ and $\eta_1$.

\section{SE Yield Estimation with SE Counting}
\label{sec:estimation}
In this section, we consider the estimation of SE yield $\eta$ under the idealized model of PBM in which SE counts are available
as described in \Cref{sec:meas-models}.
This was also a central problem of~\cite{PengMBG:21}.
Here, we provide a new interpretation of a decomposition of the Fisher information,
which we will relate to estimation with degraded observations in \Cref{sec:saturated-SEs}.
We also introduce a new estimator and provide insight on the estimators analyzed in~\cite{PengMBG:21}.

\subsection{Fisher Information in Time-Resolved Measurements}
\label{sec:FI}

The Fisher information about $\eta$ in the time resolved measurement \eqref{eq:CTTR-observation} with $\lambda$ as a known parameter was derived in~\cite{PengMBG:21}.
Normalized by $\lambda$, it is
\begin{equation}
\label{eq:CTTR-FI}
    \frac{1}{\lambda} \mathcal{I}_{\Mtilde,\Ttilde,\Xtilde}(\eta;\lambda) = \frac{1}{\eta} - e^{-\eta},
\end{equation}
which is the same as the low-$\lambda$ limit of the Fisher information in \eqref{eq:FI_low_lambda}, showing that the time-resolved measurement achieves the gain in Fisher information at low $\lambda$.

We can gain further insight into the Fisher information
by considering the contributions to it.
As was in intermediate step in deriving \eqref{eq:CTTR-FI} in~\cite{PengMBG:21},
\begin{equation}
\mathcal{I}_{\Mtilde,\Ttilde,\Xtilde}(\eta;\lambda)
= \mathcal{I}_{\Mtilde}(\eta;\lambda)  
+ \mathcal{I}_{\Xtilde|\Mtilde}(\eta;\lambda).
\end{equation}

\begin{figure}
    \centering
    \includegraphics[width=0.45\textwidth]{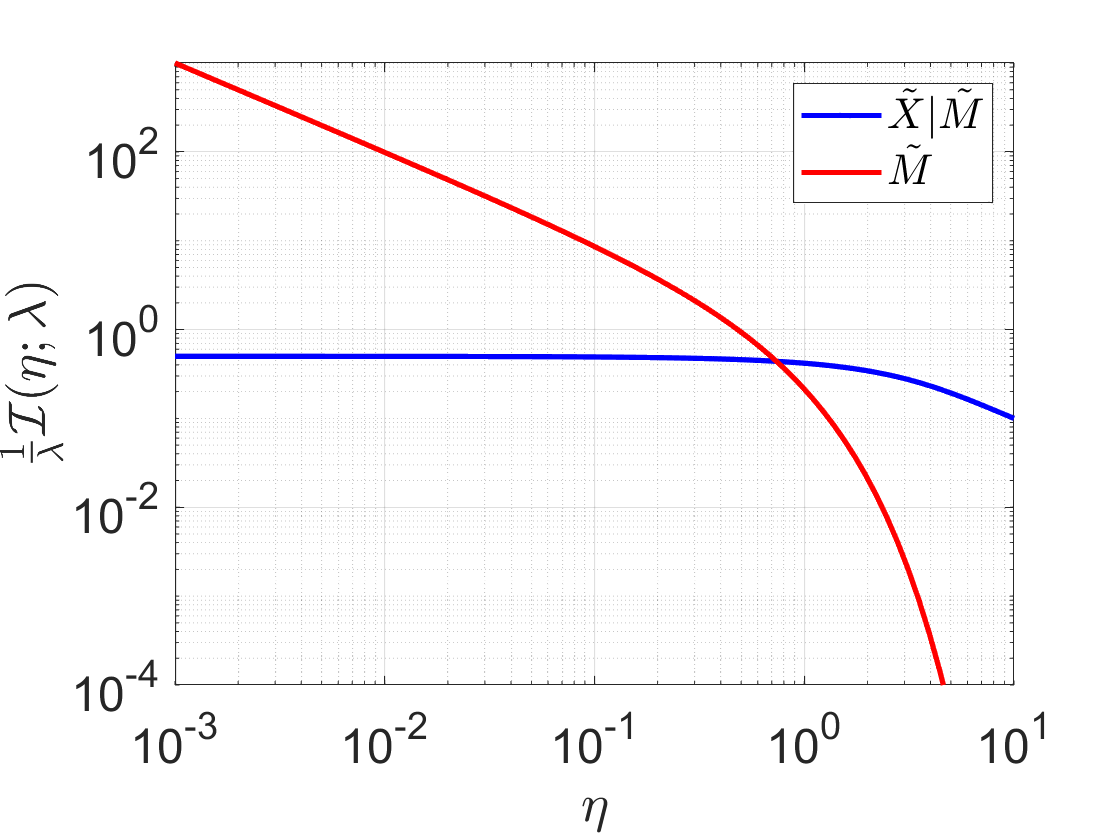}
    \caption{Fisher information in $\Mtilde$ and in $\Xtilde|\Mtilde$. The Fisher information in $\Mtilde$ dominates that in $\Xtilde|\Mtilde$ at low $\eta$, and vice-versa at high $\eta$.}
    \label{fig:FI_X_vs_M}
\end{figure}

The first term
is the information
in the number of detection events $\Mtilde$,
and the second term
is the information
in the collection of SE counts $\Xtilde$.
Normalized by $\lambda$,
the component Fisher informations are given by
\begin{equation}
\label{eq:Mtilde-FI}
\frac{1}{\lambda} \mathcal{I}_{\Mtilde}(\eta;\lambda)
= \frac{e^{-\eta}}{e^{\eta}-1}  
\end{equation}
and 
\begin{equation}
\label{eq:Xtilde-FI}
\frac{1}{\lambda} \mathcal{I}_{\Xtilde|\Mtilde}(\eta;\lambda)
= \frac{\eta+1}{\eta} - \frac{1}{1-e^{-\eta}}.
\end{equation}
\Cref{fig:FI_X_vs_M} compares these two contributions. We see that at low values of $\eta$, $\mathcal{I}_{\Mtilde}$ dominates $\mathcal{I}_{\Xtilde|\Mtilde}$, and vice-versa at higher values of $\eta$. At low $\eta$ the probability of there being more than 1 SE in a single detection event is low, which results in most $\Xtilde_i$'s being 1\@. Therefore, most of the information about $\eta$ is carried by $\Mtilde$. Conversely, at higher $\eta$, $\Mtilde \approx M$ since almost all incident particles lead to at least one detected SE\@. Therefore, information about $\eta$ is carried almost entirely by $\Xtilde$.

The observation that information about $\eta$ is available in just the number of detection events means that we can hope to estimate $\eta$ even when the measurement is saturated, \textit{i.e.}, no distinction is made between different numbers of SEs per detection. This ability to estimate $\eta$ without reference to the number of SEs in each event becomes important when the number of SEs is uncertain. In \Cref{sec:saturated-SEs}, we will introduce an $\Mtilde$-based estimator for $\eta$, and in \Cref{sec:ppg}, we will use it to estimate model parameters in a realistic PBM scenario, where noise from the SE detection chain prevents a clear distinction between the signal produced by different numbers of SEs.

\subsection{SE Yield Estimation}
\label{sec:se-count-estimators}
Prior work~\cite{PengMBG:21} introduces the following estimators for $\eta$ that can be computed from idealized TR observations:
\begin{itemize}
    \item \emph{Conventional estimator}: The conventional measurement~\eqref{eq:Y-def}, scaled by the dose $\lambda$:
    \begin{equation}
        \label{eq:eta_conv_conti}
        \etaconv = \frac{Y}{\lambda}.
    \end{equation}
    \item \emph{Oracle estimator}:
    This estimator uses the count of incident ions $M$ to improve upon $\etaconv$:
        \begin{equation}
        \label{eq:eta_conv_oracle}
        \etaoracle = \frac{Y}{M}.
    \end{equation}
    Although this estimator is clearly the best possible estimate of $\eta$, it cannot be implemented in practice since $M$ is unobservable. However, $\etaoracle$ gives useful lower bounds on the performance of TR estimators.
    \item \emph{Quotient mode (QM) estimator}:
    \begin{equation}
        \label{eq:eta_QM_conti}
        \etaQM(\Mtilde,Y) = \openCaserl
                                0, & \Mtilde = 0; \\
                \Frac{Y}{\Mtilde}, & \Mtilde > 0. \closeCase
    \end{equation}
    \item \emph{Lambert quotient mode (LQM) estimator}:
    The unique root of
    \begin{subequations}
    \label{eq:LQM}
    \begin{equation}\label{eq:LQM-fixed-point}
        \widehat{\eta} = \frac{Y}{(1-e^{-\widehat{\eta}})^{-1}\Mtilde},
    \end{equation}
    which is
    \begin{equation}\label{eq:eta_LQM_conti}
        \etaLQM = W(-\etaQM e^{-\etaQM}) + \etaQM,
    \end{equation}
    \end{subequations}
    where $W(\cdot)$ is the Lambert W function~\cite{CorlessGHJK:96}.
    \item \emph{Maximum likelihood (ML) estimator}:
    The unique root of 
    \begin{align}\label{eq:eta_trml_conti}
         \etaML = \frac{Y}{\Mtilde + \lambda e^{-\etaML}}.
    \end{align}
\end{itemize}

In~\cite{PengMBG:21}, the superior performance of the TR estimators compared to the conventional estimator was demonstrated. Before introducing a new estimator, we reinterpret these estimators,
which provides some new insight into their relative performances.

\subsubsection{LQM estimator as a mismatched ML estimate}
In~\cite{PengMBG:21}, the LQM estimator has a heuristic justification
detailed below in \Cref{sec:estimation-of-M}.
We show in this subsection that it also arises as an ML-like estimate of $\eta$
using a likelihood expression that omits the distribution of $\Mtilde$.

Suppose that $\Mtilde = \mtilde$ and $\Xtilde = (\xtilde_1,\xtilde_2,\ldots,\xtilde_{\mtilde})$
are observed.
Using the distribution of $\Xtilde_i$ from \eqref{eq:Xtilde-distribution},
the conditional likelihood of the observation given $\Mtilde = \mtilde$ is
\begin{equation}
\label{eq:LQM_mismatched_likelihood}
    \prod_{i=1}^{\mtilde}\mathrm{P}_{\Xtilde_i}(\xtilde_i \sMid \eta)
    = \left(\frac{e^{-\eta}}{1-e^{-\eta}}\right)^{\!\mtilde}
    \frac{\eta^{\xtilde_1 + \xtilde_2 + \cdots + \xtilde_{\mtilde}}}
         {\xtilde_1! \, \xtilde_2! \, \cdots \, j_{\mtilde}!}.
\end{equation}
By dropping factors that do not depend on $\eta$,
the ML-like estimate
based on \eqref{eq:LQM_mismatched_likelihood}
is
\begin{equation}
    \argmax_{\eta}\left(\frac{e^{-\eta}}{1-e^{-\eta}}\right)^{\mtilde}\eta^y,
\end{equation}

where $y = \xtilde_1 + \xtilde_2 + \cdots + \xtilde_{\mtilde}$.
The unique maximizer satisfies
\begin{equation}
    \eta = \frac{Y}{\Mtilde(1-e^{-\eta})^{-1}},
\end{equation}
which is the LQM estimator.

Note that \eqref{eq:LQM_mismatched_likelihood} is not the likelihood of the observation
$(\mtilde,\xtilde)$
because it omits the factor $\mathrm{P}_{\Mtilde}(\mtilde \sMid \eta)$.
Viewing the LQM estimator as one that ignores
the information about $\eta$ present in
$\Mtilde$ is consistent
with its generally worse performance than the ML estimator.
It is also consistent with relatively poor performance for low $\eta$,
since \Cref{fig:FI_X_vs_M} shows that $\Mtilde$ contains much more information than $\Xtilde$ for low $\eta$.

\subsubsection{Estimation of $M$}
\label{sec:estimation-of-M}
Suppose that the number of incident ions is some known positive number $m$.
Then $Y \sim \Poisson{m\eta}$,
and $\widehat{\eta} = Y/m$ is plainly the good estimator.
It is unbiased, efficient, and the ML estimate.
The number of incident ions $M$ is not directly observed,
and any information about $M$ is contained in $\Mtilde$;
conditioned on $\Mtilde$, the distributions of $\Ttilde$ and $\Xtilde$ are unrelated to $M$.

In~\cite{PengMBG:21}, the QM estimator is introduced based on plugging in $\Mtilde$ for $M$,
and the LQM estimator is introduced based on
$(1-e^{-\widehat{\eta}})^{-1}\Mtilde$
being an ad hoc improved estimate of $M$.
Specifically, since $\Mtilde \sim \binomial{M}{1-e^{-\eta}}$,
it follows that $\iE{\Mtilde \smid M} = (1-e^{-\eta})M$.
However, this does not imply that
$\iE{M \smid \Mtilde} = (1-e^{-\eta})^{-1}\Mtilde$.

In fact, there is a simple expression for $\iE{M \smid \Mtilde}$.
Using that the conditional distribution of $\Mtilde$ given $M$ is binomial
and the distribution of $M$ is $\Poisson{\lambda}$,
the conditional distribution of $M$ given $\Mtilde$ can be determined with Bayes's rule to be
\begin{equation}
\label{eq:M-given-Mtilde}
    \mathrm{P}_{M|\Mtilde}(m \smid \mtilde \sMid \eta,\lambda)
    = \frac{
    \exp(-\lambda e^{-\eta})
    (\lambda e^{-\eta})^{m-\mtilde}
    }
           {(m-\mtilde)!}, 
    \quad
\end{equation}
$m = \mtilde,\, \mtilde+1,\, \ldots$.
This is a $\Poisson{\lambda e^{-\eta}}$ distribution shifted by $\mtilde$, so
\begin{equation}
  \label{eq:BLS-of-M}
  \iE{M \smid \Mtilde} = \Mtilde + \lambda e^{-\eta}.
\end{equation}
Using this as a proxy for $M$ gives the ML estimator \eqref{eq:eta_trml_conti},
which in~\cite{PengMBG:21} is derived from maximization of the likelihood.
Putting the estimators \eqref{eq:eta_QM_conti}--\eqref{eq:eta_trml_conti}
in a single family with different proxies for $M$
explains the generally (but not uniformly) best performance of $\etaML$ and
worst performance of $\etaQM$.

\subsubsection{Conditional expectation estimator for $\eta$}
We can also use the conditional distribution of $M$ given $\Mtilde$ to develop a new estimator for $\eta$.
We have asserted that the oracle estimator $Y/M$ is a good estimate of $\eta$.
Upon observing $Y$ and $\Mtilde$, 
we can compute 
\begin{align}
    \etaCE(y,\mtilde)
      &=      \E{ Y/M         \,\big|\, Y=y,\,  \Mtilde=\mtilde } \nonumber \\
      &= y \, \E{ \frac{1}{M} \,\Big|\, \Mtilde=\mtilde}.
\label{eq:etaCE}
\end{align}

The conditional expectation is under the conditional PMF \eqref{eq:M-given-Mtilde},
yielding
\begin{equation}
    \etaCE(y,\mtilde)
       = y
       \exp(-\lambda e^{-\etaCE})
       \sum_{\ell=0}^{\infty} \frac{1}{\ell+\mtilde}
            \frac{(\lambda e^{-\etaCE})^\ell}{\ell!},
\end{equation}
which can be solved through a suitable root-finding algorithm.

\begin{figure}
  \begin{center}
    \begin{tabular}{c}
      \includegraphics[width=0.9\linewidth]{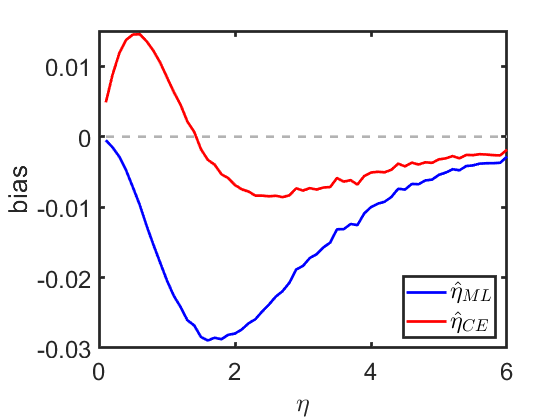} \\
      {\small (a) Bias} \\
      \includegraphics[width=0.9\linewidth]{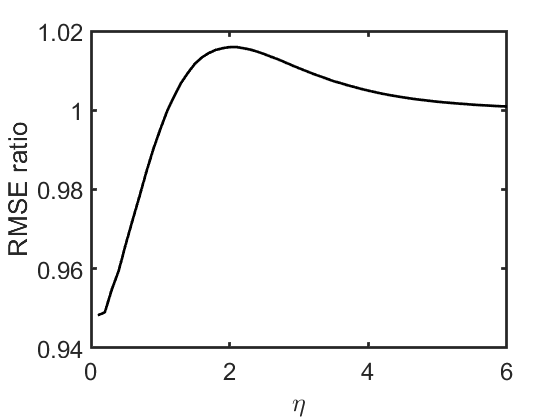} \\
      {\small (b) Root-mean squared error ratio
      }
    \end{tabular}
  \end{center}
  \caption{Numerical comparison of the $\etaCE$ estimator \eqref{eq:etaCE} with the ML estimator $\etaML$ \eqref{eq:eta_trml_conti}. $\etaCE$ has a lower magnitude of bias than $\etaML$. The RMSE ratio is close to 1 for the whole range of $\eta$.}
  \label{fig:1overM}
\end{figure}

\Cref{fig:1overM}(a) compares the bias of $\etaCE$ with that of $\etaML$, obtained using a Monte Carlo simulation for $\lambda = 20$.
We see that the magnitude of the bias of $\etaCE$ is lower than that of $\etaML$ over a wide range of $\eta$.
The variances of the two estimators are almost identical so they are not plotted.
In \Cref{fig:1overM}(b), we plot the ratio of the root mean-squared error (RMSE) of $\etaML$ to that of $\etaCE$.

\section{SE Yield Estimation from Saturated SE counts}
\label{sec:saturated-SEs}

As discussed in \Cref{sec:FI}, $\Mtilde$ contains information about $\eta$, and we can form an estimator for $\eta$ from just $\Mtilde$.
Such an estimator would treat detections as binary or saturated, since it would only consider their presence
($\Xtilde_i \geq 1$)
or absence without reference to the exact number of SEs $\Xtilde_i$ in a detection event. 

Recall from \eqref{eq:Mtilde-distribution} that
the number of incident particles that result in at least one detected SE
is given by $\Mtilde\sim\Poisson{\lambda(1-e^{-\eta})}$.
Since $\lambda$ is known, an estimator for $\eta$ would be equivalent to estimating the mean of this Poisson distribution, the ML estimate of which is the observation $\mtilde$.
When $\Mtilde < \lambda$,
we get
\begin{equation*}
    \widehat{\eta} = -\log\!\left(1-\frac{\Mtilde}{\lambda}\right)
\end{equation*}
as the ML estimate of $\eta$ from $\Mtilde$;
when $\Mtilde \geq \lambda$,
the likelihood is an increasing function of $\eta$, suggesting $\widehat{\eta} = \infty$.
Therefore, we define the estimator
\begin{equation}
    \widehat{\eta}_{\Mtilde}
        = \openCaserl
           -\log\!\left(1-\Frac{\Mtilde}{\lambda}\right), & \Mtilde < \lambda; \\
                                               \etaMax, & \Mtilde \geq \lambda, \closeCase
    \label{eq:eta_mtilde}
\end{equation}
where $\etaMax$ is some fixed value such as the largest plausible SE yield.
In the absence of an \emph{a priori} range for $\eta$, one could set $\etaMax$ to be the
largest possible value returned when $\Mtilde < \lambda$:
\begin{equation}
    \etaMax = -\log\!\left( 1 - \frac{\lceil \lambda \rceil - 1}{\lambda} \right).
\end{equation}
However, this expression has large jumps at integer values of $\lambda$.
The choice of
\begin{equation}
\label{eq:etaMax-rule}
    \etaMax = -\log\!\left( 1 - \frac{\lceil \lambda \rceil - 1}{\lceil \lambda \rceil} \right)
\end{equation}
is more conservative.

\begin{figure}
  \begin{center}
    \begin{tabular}{c}
      \includegraphics[width=0.9\linewidth]{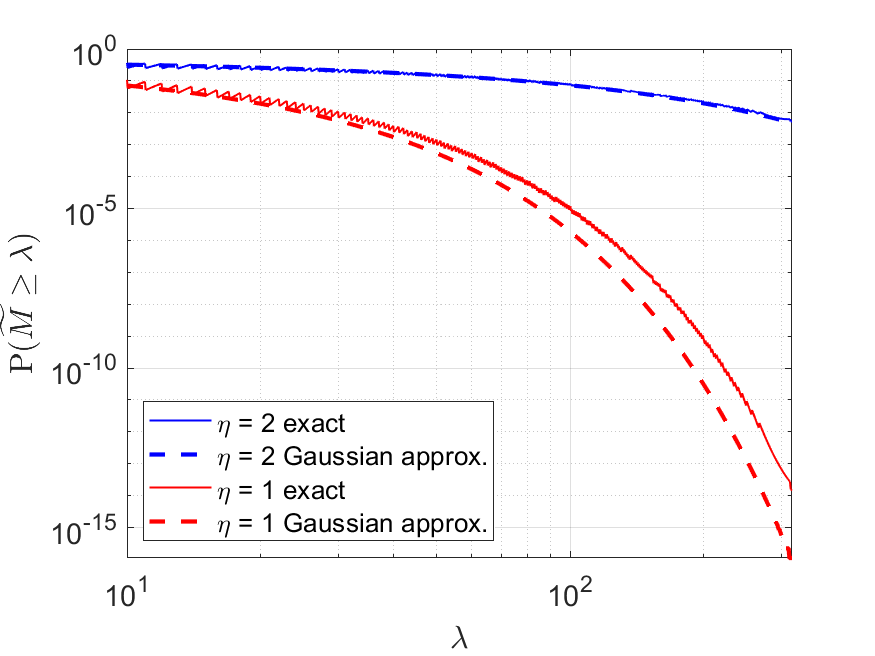} \\
      {\small (a) Dependence on $\lambda$ with $\eta$ fixed} \\
      \includegraphics[width=0.9\linewidth]{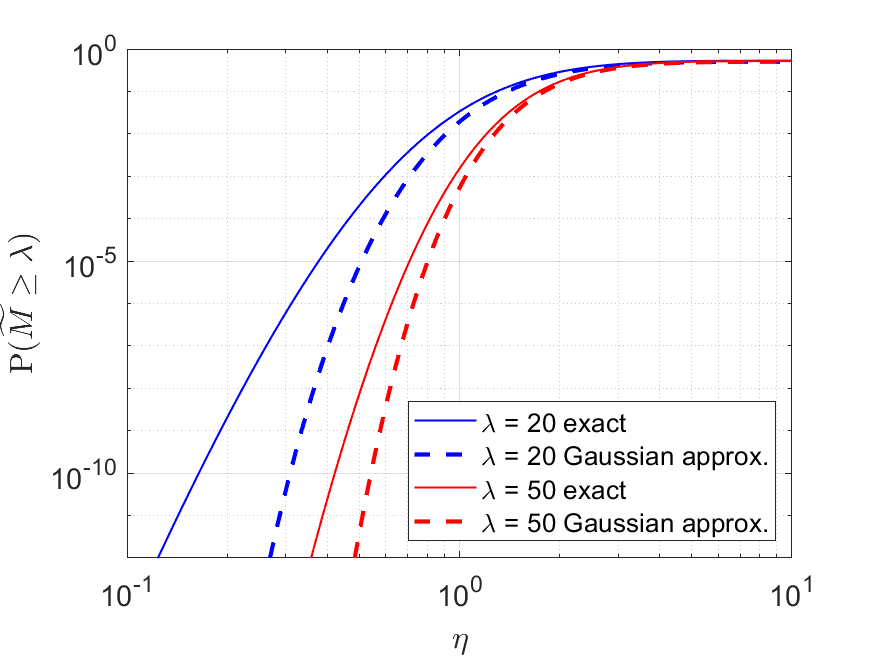} \\
      {\small (b) Dependence on $\eta$ with $\lambda$ fixed}
    \end{tabular}
  \end{center}
  \caption{$\Mtilde$-based estimator for $\eta$ fails when $\iP{\Mtilde \geq \lambda}$.
    This occurs with vanishing probability as (a) $\lambda$ increases or as (b) $\eta$ decreases.}
  \label{fig:Mtilde_estimator_fails}
\end{figure}

For any fixed $\eta$, the probability of $\Mtilde \geq \lambda$ decreases with increasing $\lambda$;
for any fixed $\lambda$, the probability of $\Mtilde \geq \lambda$ decreases with decreasing $\eta$.
These trends are illustrated in \Cref{fig:Mtilde_estimator_fails}.
Typical values of $\lambda$ for imaging range from $\sim 10$ to $\sim 100$.
However, as described in \Cref{sec:estimating-ppg-parameters},
much larger values may arise in calibration.

\begin{figure*}
  \begin{center}
    \begin{tabular}{@{}c@{\,}c@{\,}c@{}}
      \includegraphics[width=0.33\linewidth]{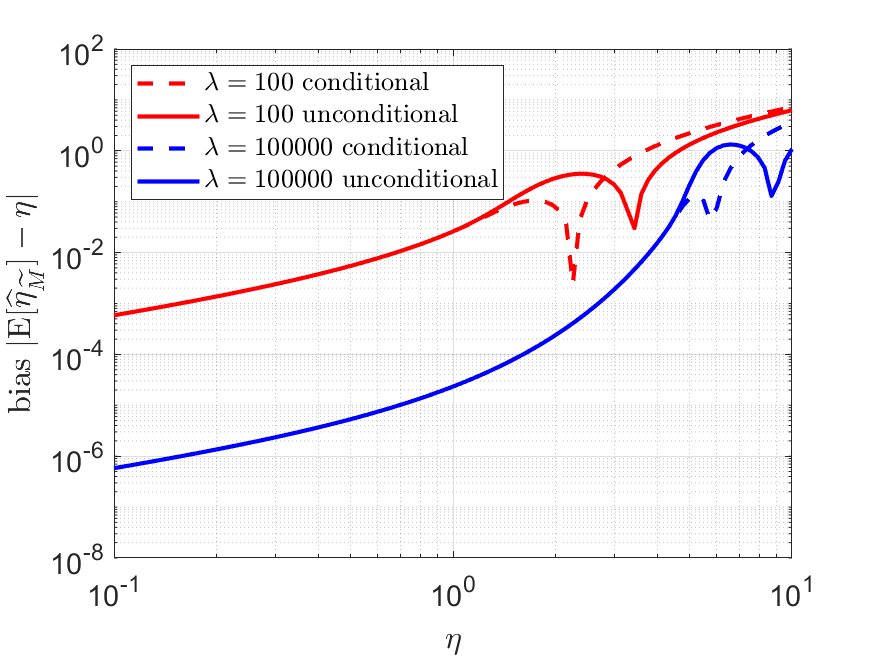} &
      \includegraphics[width=0.33\linewidth]{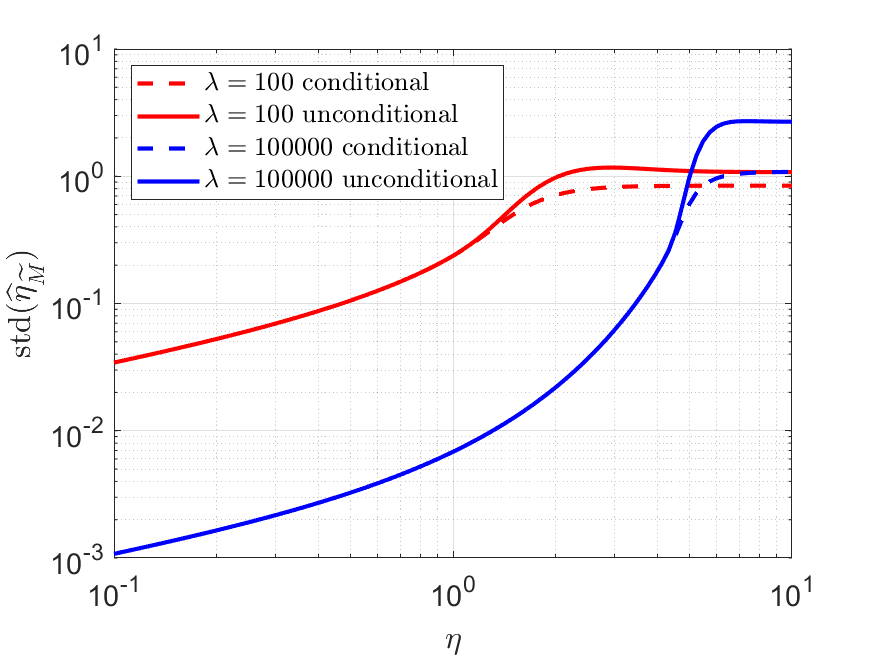} &
      \includegraphics[width=0.33\linewidth]{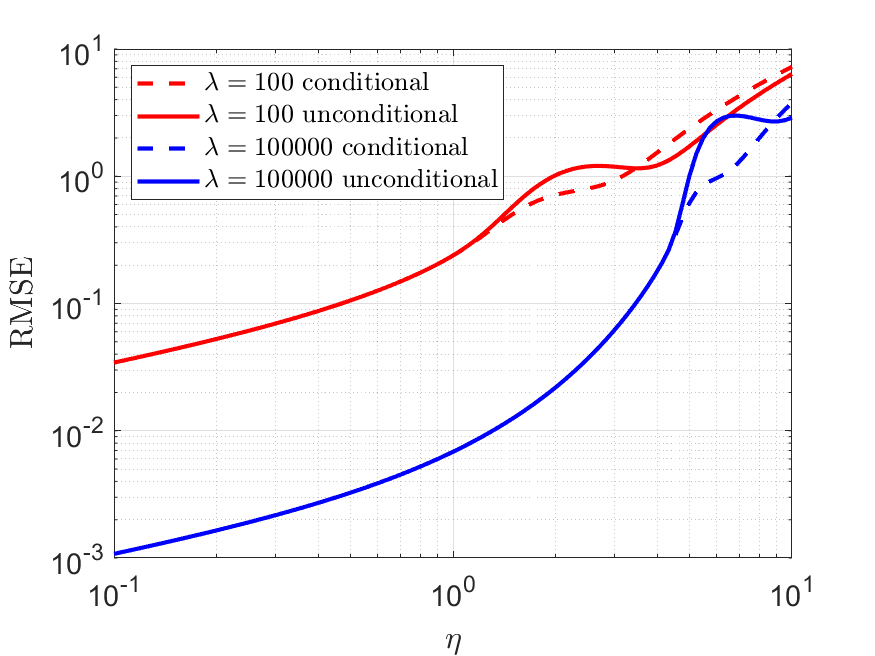} \\
      {\small (a) Bias} &
      {\small (b) Standard deviation} &
      {\small (c) Root mean-squared error}
    \end{tabular}
  \end{center}
    \caption{Performance of $\etaMtilde$, the estimator \eqref{eq:eta_mtilde} that uses the number of SE detection events without SE count information.
    Conditional curves give the indicated quantity conditioned on
    $\Mtilde < \lambda$, which is when the ML estimate is finite.
    Unconditional curves give the indicated quantity including the effect
    of choosing $\etaMax$ according to \eqref{eq:etaMax-rule}.
    }
    \label{fig:Mtilde_estimator_performance}
\end{figure*}

The performance of the estimator is shown in \Cref{fig:Mtilde_estimator_performance}
for $\lambda = 100$ and $\lambda = 100\,000$,
where $\etaMax$ is set using \eqref{eq:etaMax-rule}
and $\eta \in [\frac{1}{10},\,10]$.
Bias and variance are separated,
and we can see that the RMSE
is dominated by variance at low $\eta$ and by bias at high $\eta$.
The variance levels off around half of $\etaMax$.
Kinks in the absolute bias curves are due to the bias changing sign from positive for low $\eta$
to negative for high $\eta$.

From our analysis of the Fisher information about $\eta$ in $\Mtilde$ in \Cref{fig:FI_X_vs_M},
we expect that this estimator gets worse as $\eta$ increases, which is indeed what we observe.
In the next section, we will use $\etaMtilde$ to estimate the parameters for a PBM model that includes uncertainty in SE number introduced by the SE detection chain.

\section{SE Yield Estimation from SE Detector Voltages}
\label{sec:ppg}
As  described in \Cref{sec:typical-instrument}, in a real particle beam microscope, direct counts of secondary electrons are usually not available. Instead, as depicted in \Cref{fig:marked-process-illustration}(e), the output of the SE detector is a series of voltage pulses. Assuming that the conversion of SE number to voltage pulse is linear, we expect that the average height of each pulse is proportional
to the count of SEs incident on the detector. When $\eta$ is low, the probability that an incident particle generated multiple SEs is low, and a count of pulses can be used to estimate the true SE count. This scenario is true in SEM, and pulse counting has been used to implement SE count imaging in SEM~\cite{Yamada1990, Yamada1990a, Yamada1991, Yamada1991a, Uchikawa1992, AGARWAL2021, Agarwal:2023-U}. However, if $\eta$ is higher, as in HIM, excitation of multiple SEs by a single incident particle becomes more likely. Therefore, more sophisticated modelling is needed to estimate $\eta$.

In this section, we will describe a probabilistic model for the observed SE voltage signal and analyze how the Fisher information about $\eta$ varies with model parameters. We will also discuss how the model parameters may be estimated. Finally, we will discuss the performance of $\eta$ estimators based on this model.

\subsection{Pulse Height Model}
As in~\cite{PengMBBG:20}, we model the conversion of SE number to voltages with a \textit{Poisson--Poisson--Gaussian} (PPG) model.
Each SE is assumed to produce a voltage described by a $\normal{c_1}{c_2}$ random variable, where $c_1$ is the mean voltage and $c_2$ the variance.
These contributions are assumed to be independent and additive, so $j$ SEs produce a voltage with the $\normal{jc_1}{jc_2}$ distribution.
Thus, the probability density function for the voltage $\Utilde_i$
produced in the $i$th detection event is given by

\begin{align}
    f_{\Utilde_i}( u \sMid \eta,c_1,c_2)
    & = \sum_{j=1}^{\infty} \mathrm{P}_{\Xtilde}(j;\eta)
    f_Z(u \sMid j,c_1,c_2) \nonumber \\
    & = \sum_{j=1}^{\infty} \frac{e^{-\eta}}{1-e^{-\eta}}\frac{\eta^j}{j!}
    f_Z(u \sMid j,c_1,c_2),
    \label{eq:ppg_pdf}
\end{align}
where $f_Z(u \sMid j,c_1,c_2)$ is the PDF of a $\normal{jc_1}{jc_2}$ random variable.

The heights of the detected SE pulses, along with the total number of pulses, form the time-resolved observation
\begin{align}
\left\{ \Mtilde, \Ttilde, \Utilde \right\} 
    &= \left\{ \Mtilde,
       (\Ttilde_1,\Ttilde_2,...,\Ttilde_{\Mtilde}),
       (\Utilde_1,\Utilde_2,...,\Utilde_{\Mtilde}) \right\}.
\label{eq:PPG-observation}
\end{align}
Under this model, a conventional observation without time resolution is
\begin{equation}
\label{eq:V-def}
    V = \sum_{i=1}^{\Mtilde} \Utilde_i.
\end{equation}
Analogously to the discussion in \Cref{sec:tr-model},
conditioned on $\Mtilde$,
there is no information about $\eta$ in $\Ttilde$.

\subsection{Fisher Information}

\begin{figure}
    \centering
    \includegraphics[width=0.45\textwidth]{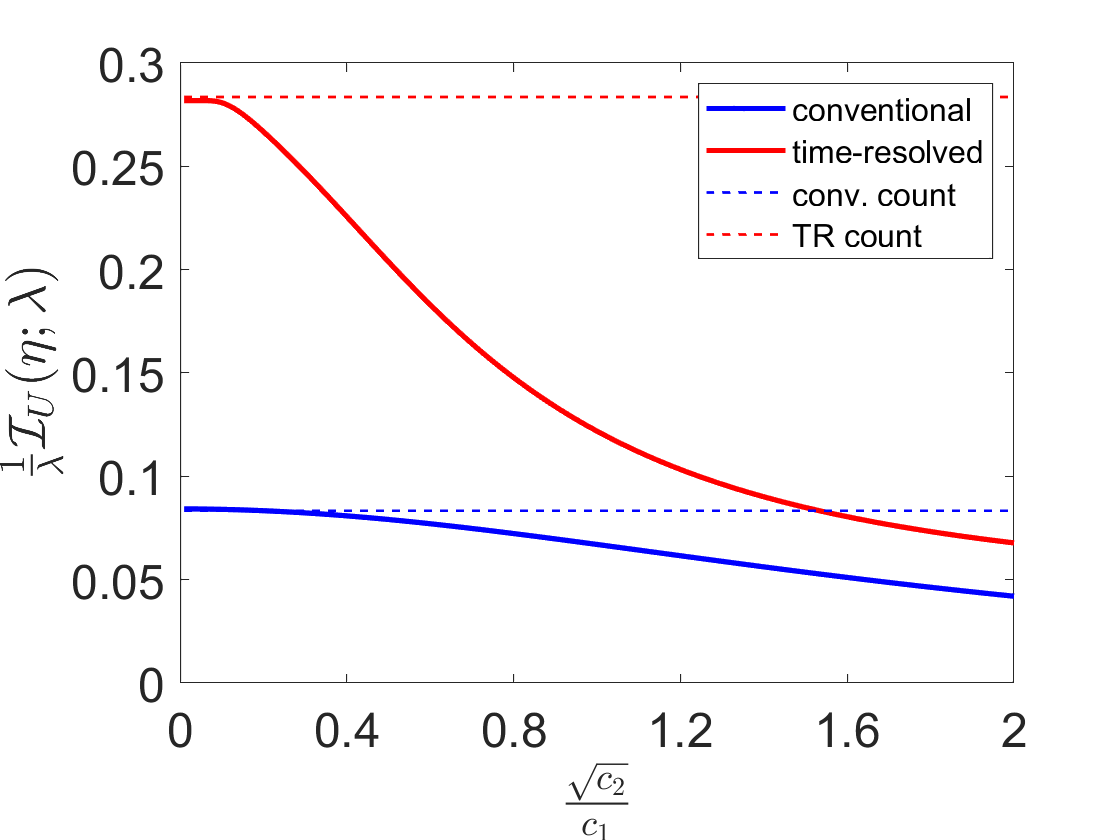}
    \caption{Normalized Fisher information
    $\frac{1}{\lambda}\mathcal{I}_{\Mtilde,\Utilde}(\eta;\lambda)$
    for time-resolved measurement
    and 
    $\frac{1}{\lambda}\mathcal{I}_{V}(\eta;\lambda)$
    for conventional measurement
    as a function of $\Frac{\sqrt{c_2}}{c_1}$ for $\lambda = 20$ and $\eta = 3$.}
    \label{fig:fi_ppg}
\end{figure}

We can evaluate the Fisher information numerically.
\Cref{fig:fi_ppg} is a plot of the Fisher information (normalized by $\lambda$) for the PPG model,
for both conventional and time-resolved measurements,
as a function of $\Frac{\sqrt{c_2}}{c_1}$, at $\eta = 3$.
For the time-resolved case, when $\Frac{\sqrt{c_2}}{c_1} \leq 0.1$, the FI is nearly constant.
At such low values of $\Frac{\sqrt{c_2}}{c_1}$,
there is little overlap between the peaks in the probability density of $\Utilde_i$ produced by different numbers of SEs, resulting in near-perfect discrimination of SE counts.
Thus, the FI reaches the marked asymptote, which is the FI with SE counting \eqref{eq:CTTR-FI}.
Similarly, the FI for conventional measurement reaches the asymptote \eqref{eq:FI_high_lambda}.
As $\Frac{\sqrt{c_2}}{c_1}$ increases, the FI degrades, reflecting the ambiguity in resolving the number of SEs due to overlap in the densities of different numbers of SEs.
We note that although the FI from time-resolved measurements remains higher than that for conventional measurements for the whole range of $\Frac{\sqrt{c_2}}{c_1}$, the relative advantage of time-resolved measurements diminishes at higher values of $\Frac{\sqrt{c_2}}{c_1}$.

\subsection{Estimating $c_1$ and $c_2$}
\label{sec:estimating-ppg-parameters}
The PPG model parameters, $c_1$ and $c_2$, are generally unknown for a given particle-beam microscope.
The values of these parameters depend on the SE detector hardware settings, such as the gain in the dynode stages of the photomultiplier tube, the specifics of the pre-amplifier circuit, etc.
The values of these settings are unavailable to the user.
Therefore the model parameters cannot be directly computed. 

Instead, we must estimate the parameters $c_1$ and $c_2$. We could conveniently do so
if we image a sample with a well-characterized $\eta$.
In this case, we could find an ML estimate for $c_1$ and $c_2$ by maximizing the likelihood of the observed pulse heights under the PPG model in \eqref{eq:ppg_pdf}.
Although standard values of $\eta$ for different materials are widely available~\cite{Seiler1983, Joy2006}, the precise value of $\eta$ for a given sample depends on several factors such as the level of carbon contamination in the microscope vacuum chamber and surface oxidation, making estimation of $c_1$ and $c_2$ difficult.
However, we can use $\etaMtilde$ given in \eqref{eq:eta_mtilde} to estimate $\eta$, since this estimator does not rely on the number of SEs
(\textit{i.e.,} the heights of the detected voltage pulses), but only on the number of detected pulses.
Therefore, it is not affected by the loss in Fisher information due to variance in pulse heights depicted in \Cref{fig:fi_ppg}.
With this $\eta$ estimate, we can construct ML estimates for $c_1$ and $c_2$. 

\begin{figure}
    \centering
    \includegraphics[width=0.45\textwidth]{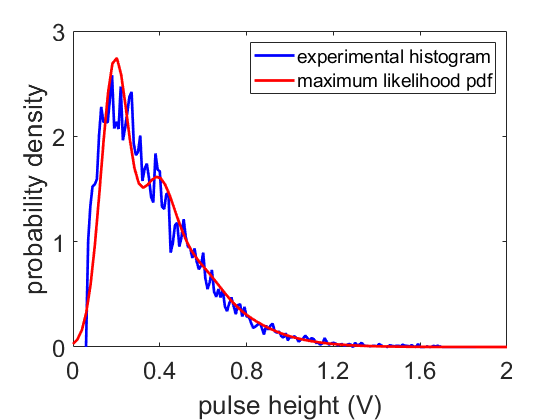}
    \caption{Estimation of PPG model parameters $c_1$ and $c_2$. The probability distribution with the ML estimates of the model parameters is plotted along with the observed pulse height histogram.}
    \label{fig:ppg_parameter_estimation}
\end{figure}

To demonstrate this process, we imaged a uniform, featureless silicon chip on an HIM (Zeiss Orion) at a resolution of $10^5$ pixels with a beam current of 0.1 $\pA$ and a pixel dwell time of 10 $\us$, which corresponds to $\lambda = 6.25$.
\Cref{fig:marked-process-illustration}(e) shows a snapshot of the voltage pulses detected from one pixel in the image.
The average $\Mtilde$ (per pixel) observed for this sample was 4.95\@.
Since the entire sample was treated as uniform,
we used the sum of $\Mtilde$ over all the pixels and the total $\lambda$ over all the pixels in \eqref{eq:eta_mtilde}
to obtain $\etaMtilde = 1.58$.
Next, we used this estimate of $\eta$ to construct ML estimates of $c_1$ and $c_2$.
The resulting probability density function is shown in \Cref{fig:ppg_parameter_estimation}, along with the distribution of pulse heights in the experimental data. The ML estimates of $c_1$ and $c_2$ obtained from this technique were $c_1=0.19~\V$ and $c_2=0.0040~\V^2$.

We end this section with a couple of observations about the estimation of $c_1$ and $c_2$ using $\etaMtilde$.
First, although we can use $\etaMtilde$ in the estimation of $c_1$ and $c_2$, we will not use it to estimate $\eta$ during imaging.
As discussed in \Cref{sec:saturated-SEs} and shown in \Cref{fig:Mtilde_estimator_performance},
for good accuracy this estimator requires the dose to be very high.
Second, even though the model PDF shows a good fit with the experimental pulse height histogram, it predicts a significant density of pulses with heights near and below zero volts. This is clearly unphysical, and it points to a mismatch between the model and the experiment.

\subsection{Accounting for Nonzero Pulse Widths} 

An additional feature of the voltage pulses is a nonzero width in the time domain.
Empirically,
from experimental pulse sequences such as that in \Cref{fig:marked-process-illustration}(e),
we found the pulses to be approximately Gaussian in shape with mean width of
$\tau = 160\,\ns$.
The nonzero
widths
raise the possibility of undercounting SE detection events due to overlap between detections from successive incident particles.
To compensate for this undercounting, we introduce a correction factor $\gamma_\tau(\lambda,\eta)$ such that
\begin{equation}
    \Mtildecorr = \frac{\Mtilde}{\gamma_\tau(\lambda,\eta)}.
\end{equation}
This correction factor is obtained by integrating the exponential probability distribution of the SE interarrival times up to
the mean pulse width $\tau$.
This gives us
\begin{equation}
    \gamma_\tau(\lambda,\eta) = \exp\!\left(-\lambda(1-e^{-\eta}\right)\tau).
\end{equation}
Some of the estimators described below use $\Mtilde_{\textrm{corrected}}$.

\subsection{Estimators}
\label{subsec:ppg-estimators}
\begin{figure*}
  \begin{center}
    \begin{tabular}{c@{}c@{}c@{}}
      \includegraphics[width=0.33\linewidth]{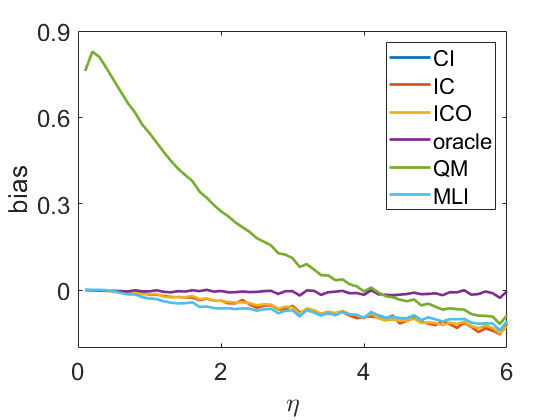} &
      \includegraphics[width=0.33\linewidth]{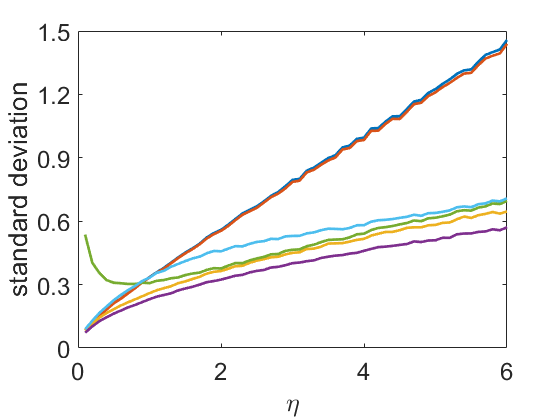} &
      \includegraphics[width=0.33\linewidth]{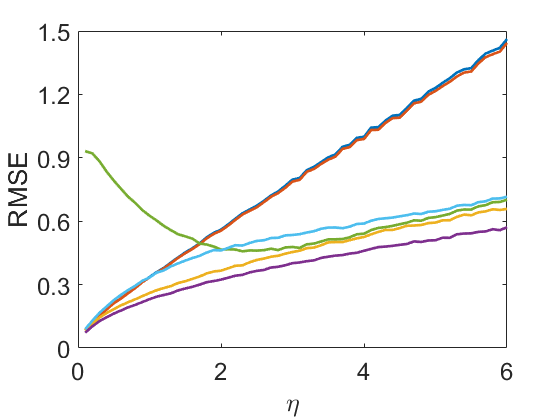} \\
      {\small (a) Bias} &
      {\small (b) Standard deviation} &
      {\small (c) Root mean-squared error} 
    \end{tabular}
  \end{center}
  \caption{Relative performance of PPG estimators, for $c_1 = 0.19~\V$, $c_2 = 0.011~\V^2$, and $\lambda = 20$. $\etaMLI$ matches the performance of $\etaCI$ and $\etaIC$ at low $\eta$, and significantly outperforms them for $\eta>1$.}
  \label{fig:ppg_estimators}
\end{figure*}

We now introduce $\eta$ estimators suitable for the PPG model.
We construct these estimators to be analogous to estimators using SE counts from \Cref{sec:se-count-estimators}.
For comparison, we include a model for a typical instrument, a somewhat idealized conventional estimator, and two oracles.

\subsubsection{Conventional instrument}
As described in \Cref{sec:typical-instrument}, a typical instrument forms an image by sampling the voltage output from the SE detector, adding up the samples for each pixel, and quantizing these summed values to obtain an 8-bit image.
For the purpose of comparing with other $\eta$ estimators, we define $\etaCI$ based on emulating this process.
For each incident particle, a pulse with height following \eqref{eq:ppg_pdf} and width $\tau$ is generated.
The resulting waveform is sampled with period 100\,$\ns$ and summed to obtain an estimate.
An additional factor is needed to be on the correct scale;
we determine this scaling factor by matching the mean to the mean of the improved conventional estimate below.

\subsubsection{Improved conventional}
Within the PPG model, $V$ in \eqref{eq:V-def} contains all the information acquirable without time resolution.
Dividing by $c_1$ puts $V$ on the scale of SE yield $\eta$,
so analogously to \eqref{eq:eta_conv_conti} we define
\begin{equation}
    \etaIC = \frac{V/c_1}{\lambda}.
\end{equation}

\subsubsection{Ion count oracle}
To contextualize the performance of the implementable estimators,
we use the oracle from \eqref{eq:eta_conv_oracle} along with
an ion count oracle that assumes the true count of ions $M$ is known:
\begin{equation}
    \etaICO = \frac{V/c_1}{M}.
\end{equation}

\subsubsection{QM estimator}
Analogous to the QM estimator in \Cref{sec:se-count-estimators}, we can use $\Mtildecorr$ as a proxy for $M$.
Then our estimate is the unique root of
\begin{equation}
    \etaQM = \frac{V/c_1}
    {\Frac{\Mtilde}{\gamma_\tau(\lambda,\etaQM)}}.
\end{equation}

\subsubsection{ML-inspired estimator}
We can correct the bias in $\Mtildecorr$ analogously to the ML estimator in \Cref{sec:se-count-estimators},
noting that this is not a true ML estimator under our current model.
The estimate is the unique root of
\begin{equation}
    \etaMLI = \frac{V/c_1}{
    {\Frac{\Mtilde}{\gamma_\tau(\lambda,\etaMLI)}}
    + \lambda e^{-\etaMLI}}.
\end{equation}

\Cref{fig:ppg_estimators} plots the bias, standard deviation, and RMSE of these estimators as functions of $\eta$, using the values of $c_1$ and $c_2$ from \Cref{sec:estimating-ppg-parameters} and $\lambda = 20$.
These values were calculated using a Monte Carlo simulation.
The most important observation from this figure is that the TR estimators continue to outperform the conventional estimator over almost the entire range of $\eta$ considered here.
Similar to the results in~\cite{PengMBG:21}, $\etaQM$ has a high bias at low $\eta$ due to significant underestimation of $M$, but its RMSE is still lower than the conventional estimators for $\eta>1.9$.
The two conventional estimators, $\etaCI$ and $\etaIC$, have almost identical performance; however, without calculation of $\etaIC$, the factor $k$ in $\etaCI$ would be unknown and the RMSE potentially larger.
The ion count oracle $\etaICO$
forms an effective lower bound on the
RMSE
of the non-oracle estimators, and $\etaQM$ and $\etaMLI$ achieve performance close to $\etaICO$ for high $\eta$.
The RMSE of $\etaoracle$ is lower than that of $\etaICO$ by a factor of $\sim1.4$, reflecting the loss in information about $\eta$ from increased uncertainty in the SE counts.

\section{Conclusion}
\label{sec:conclusion}
In this work, we have shown that TR measurements, where we measure the full vector of SE detections for every pixel, outperforms conventional scalar-valued PBM for detecting changes in $\eta$ or estimating $\eta$.
We motivated TR measurements by quantifying a gain in Fisher information for estimation of $\eta$ at low dose $\lambda$, as well as increased error exponents for discrimination between two values of $\eta$ using KLD. We also demonstrated that TR estimators outperform the conventional estimator for $\eta$ both in the idealized scenario where direct counts of detected SEs are available
(yielding the measurement vector
$\{ \Mtilde, \Ttilde, \Xtilde \}
= \{ \Mtilde, (\Ttilde_1,\Ttilde_2,\ldots,\Ttilde_{\Mtilde}),(\Xtilde_1,\Xtilde_2,\ldots,\Xtilde_{\Mtilde}) \}$),
as well as the more realistic scenario where noise from the SE detection process makes direct SE counts inaccessible
(making the measurement vector
$\{ \Mtilde, \Ttilde, \Utilde \}
= \{ \Mtilde,(\Ttilde_1,\Ttilde_2,\ldots,\Ttilde_{\Mtilde}),(\Utilde_1,\Utilde_2,\ldots,\Utilde_{\Mtilde}) \}$).

Our re-analysis of previously derived estimators for $\eta$ led to new insights into their relative performances.
We also developed two new estimators:
the conditional expectation estimator using the conditional distribution of $M$ given $\Mtilde$; and
an $\Mtilde$-based estimator that only uses the count of SE pulses.
The latter estimator was particularly useful for deriving values for the PPG model parameters $c_1$ and $c_2$. 

The estimator $\etaMtilde$ could also be used to calculate the PBM detector efficiency~\cite{Joy1996,Joy2006}.
For this application, a bulk sample with $\eta$ known under the imaging conditions being used would need to be imaged (or, alternatively, $\eta$ of the bulk sample could be measured in the PBM using a standard technique~\cite{Seiler1983}).
Then, the ratio of $\etaMtilde$ and the true sample $\eta$ would be the instrument's detective efficiency.
A similar technique was used in \cite{AGARWAL2021} to calculate detector efficiency;
that work was further simplified by the low value of $\eta$, which allows detection efficiency to be estimated as the ratio of $\Mtilde$ and $\lambda$ times the known $\eta$.

Successful implementation of the $\eta$ estimators described in \Cref{subsec:ppg-estimators} depends on the accuracy of the PPG model in representing the instrument response, as well as accurate estimation of model parameters.
As discussed previously, the PPG model leaves open the unphysical possibility of negative SE voltage pulse heights.
This issue could be resolved with a heuristic approach, such as zero-truncation or folding of the probability density function.
Alternatively, a pulse height histogram such as the one in \Cref{fig:ppg_parameter_estimation} could be acquired at a low $\eta$, such that the probability of more than one SE being detected from a given pixel is sufficiently small.
Such a histogram could then be used as the empirical single-SE instrument response.

The loss in Fisher information with increasing uncertainty in the count of SEs, as demonstrated in \Cref{fig:fi_ppg}, points to the potential benefits of hardware SE counting in PBM\@.
Although direct counting of SEs is available in transmission-based PBM techniques such as transmission electron microscopy~\cite{Ishikawa2014,Sang2016,McMullan2016}, it has not been explored in SE-based techniques.
Its implementation in SEM and HIM could lead to large improvements in achievable image quality.

\section*{Acknowledgements}
The authors acknowledge Dr. Leila Kasaei, Dr. Hussein Hijazi, and Prof. Leonard Feldman from the Department of Physics, Rutgers University for enabling the acquisition of the experimental data used in this work, as well as fruitful discussions on its interpretation.

\bibliographystyle{IEEEtran}
\bibliography{bibl, bibiligraph}

\newcommand{\SortNoop}[1]{}
\begin{thebibliography}{10}
\providecommand{\url}[1]{#1}
\csname url@samestyle\endcsname
\providecommand{\newblock}{\relax}
\providecommand{\bibinfo}[2]{#2}
\providecommand{\BIBentrySTDinterwordspacing}{\spaceskip=0pt\relax}
\providecommand{\BIBentryALTinterwordstretchfactor}{4}
\providecommand{\BIBentryALTinterwordspacing}{\spaceskip=\fontdimen2\font plus
\BIBentryALTinterwordstretchfactor\fontdimen3\font minus
  \fontdimen4\font\relax}
\providecommand{\BIBforeignlanguage}[2]{{%
\expandafter\ifx\csname l@#1\endcsname\relax
\typeout{** WARNING: IEEEtran.bst: No hyphenation pattern has been}%
\typeout{** loaded for the language `#1'. Using the pattern for}%
\typeout{** the default language instead.}%
\else
\language=\csname l@#1\endcsname
\fi
#2}}
\providecommand{\BIBdecl}{\relax}
\BIBdecl

\bibitem{Oatley:82}
C.~W. Oatley, ``The early history of the scanning electron microscope,''
  \emph{J. Appl. Physics}, vol.~53, no.~2, Feb. 1982.

\bibitem{McMullan1995}
D.~McMullan, ``Scanning electron microscopy 1928--1965,'' \emph{Scanning},
  vol.~17, no.~3, pp. 175--185, May--Jun. 1995.

\bibitem{erwin1969field}
E.~W. M{\"u}ller and T.~T. Tsong, \emph{Field Ion Microscopy: Principles and
  Applications}.\hskip 1em plus 0.5em minus 0.4em\relax American Elsevier,
  1969.

\bibitem{WardNE:06}
B.~W. Ward, J.~A. Notte, and N.~P. Economou, ``Helium ion microscope: A new
  tool for nanoscale microscopy and metrology,'' \emph{J. Vacuum Sci. Technol.
  B}, vol.~24, no.~6, pp. 2871--2874, Nov. 2006.

\bibitem{Joens:13}
M.~S. Joens, C.~Huynh, J.~M. Kasuboski, D.~Ferranti, Y.~J. Sigal, F.~Zeitvogel,
  M.~Obst, C.~J. Burkhardt, K.~P. Curran, S.~H. Chalasani, L.~A. Stern,
  B.~Goetze, and J.~A.~J. Fitzpatrick, ``Helium ion microscopy ({HIM}) for the
  imaging of biological samples at sub-nanometer resolution,'' \emph{Sci.
  Rep.}, vol.~3, no. 3514, Dec. 2013.

\bibitem{Merolli2022}
A.~Merolli, L.~Kasaei, S.~Ramasamy, A.~Kolloli, R.~Kumar, S.~Subbian, and L.~C.
  Feldman, ``An intra-cytoplasmic route for sars-cov-2 transmission unveiled by
  helium-ion microscopy,'' \emph{Scientific Reports}, vol.~12, Dec. 2022.

\bibitem{CastaldoHKVM:09}
V.~Castaldo, C.~W. Hagen, P.~Kruit, E.~van Veldhoven, and D.~Maas, ``On the
  influence of the sputtering in determining the resolution of a scanning ion
  microscope,'' \emph{J. Vacuum Sci. Technol. B}, vol.~27, no.~6, pp. 982--994,
  2009.

\bibitem{CastaldoHK:11}
V.~Castaldo, C.~W. Hagen, and P.~Kruit, ``Simulation of ion imaging:
  Sputtering, contrast, noise,'' \emph{Ultramicroscopy}, vol. 111, pp.
  982--994, 2011.

\bibitem{Dahmen2016}
T.~Dahmen, M.~Engstler, C.~Pauly, P.~Trampert, N.~de~Jonge, F.~Mücklich, and
  P.~Slusallek, ``Feature adaptive sampling for scanning electron microscopy,''
  \emph{Scientific Reports}, vol.~6, p. 25350, Jul. 2016.

\bibitem{BarlowPSC:16}
A.~J. Barlow, J.~F. Portoles, N.~Sano, and P.~J. Cumpson, ``Removing beam
  current artifacts in helium ion microscopy: A comparison of image processing
  techniques,'' \emph{Microsc. Microanal.}, vol.~22, no.~5, pp. 939--947, Oct.
  2016.

\bibitem{PengCDIKKG:21}
M.~Peng, M.~Cokbas, U.~{Dorken Gallastegi}, P.~Ishwar, J.~Konrad, B.~Kulis, and
  V.~K. Goyal, ``Convolutional neural network denoising of focused ion beam
  micrographs,'' in \emph{Proc. IEEE 31st Int. Workshop Mach. Learn. Signal
  Process.}, Gold Coast, Queensland, Australia, Oct. 2021.

\bibitem{Vanderlinde2007}
W.~E. Vanderlinde and J.~N. Caron, ``Blind deconvolution of sem images,'' in
  \emph{Proc. 33rd Int. Symp. Testing and Failure Analysis}, Nov. 2007, pp.
  97--102.

\bibitem{Pang2021}
S.~Pang, X.~Zhang, H.~Li, and Y.~Lu, ``Edge determination improvement of
  scanning electron microscope images by inpainting and anisotropic diffusion
  for measurement and analysis of microstructures,'' \emph{Measurement}, vol.
  176, p. 109217, 2021.

\bibitem{Dahmen2018}
P.~Potocek, R.~Schoenmakers, P.~Trampert, T.~Dahmen, and M.~Peemen, ``Sparse
  scanning electron microscopy for imaging and segmentation in connectomics,''
  in \emph{Proc. IEEE Int. Conf. Bioinformatics and Biomedicine (BIBM)}, 2018,
  pp. 2461--2465.

\bibitem{PengMBBG:20}
M.~Peng, J.~Murray-Bruce, K.~K. Berggren, and V.~K. Goyal, ``Source shot noise
  mitigation in focused ion beam microscopy by time-resolved measurement,''
  \emph{Ultramicroscopy}, vol. 211, no. 112948, Apr. 2020.

\bibitem{PengMBG:21}
M.~Peng, J.~Murray-Bruce, and V.~K. Goyal, ``Time-resolved focused ion beam
  microscopy: Modeling, estimation methods, and analyses,'' \emph{IEEE Trans.
  Comput. Imaging}, vol.~7, pp. 547--561, 2021.

\bibitem{Watkins2021a}
L.~Watkins, S.~W. Seidel, M.~Peng, A.~Agarwal, C.~C. Yu, and V.~K. Goyal,
  ``Robustness of time-resolved measurement to unknown and variable beam
  current in particle beam microscopy,'' in \emph{Proc. IEEE Int. Conf. Image
  Process.}, Anchorage, AK, Sep. 2021, pp. 3487--3491.

\bibitem{Watkins2021b}
------, ``Prevention beats removal: Avoiding stripe artifacts from current
  variation in particle beam microscopy through time-resolved sensing,''
  \emph{Microsc. Microanal.}, vol.~27, no.~S1, pp. 422--425, Aug. 2021.

\bibitem{Seidel2022TCI}
S.~W. Seidel, L.~Watkins, M.~Peng, A.~Agarwal, C.~Yu, and V.~K. Goyal, ``Online
  beam current estimation in particle beam microscopy,'' \emph{IEEE Trans.
  Comput. Imaging}, vol.~8, pp. 521--535, 2022.

\bibitem{Seidel2022_MM}
S.~W. Seidel, L.~Watkins, M.~Peng, A.~Agarwal, C.~C. Yu, and V.~K. Goyal,
  ``Addressing neon gas field ion source instability through online beam
  current estimation,'' \emph{Microsc. Microanal.}, vol.~28, no.~S1, pp.
  36--39, Aug. 2022.

\bibitem{PengKSYG:23arXiv}
M.~Peng, R.~Kitichotkul, S.~W. Seidel, C.~Yu, and V.~K. Goyal, ``Denoising
  particle beam micrographs with plug-and-play methods,'' arXiv:2208.14256v2
  [physics.med-ph]., Feb. 2023.

\bibitem{Everhart1960}
T.~E. Everhart and R.~F. Thornley, ``Wide-band detector for micro-microampere
  low-energy electron currents,'' \emph{J. Scientific Instruments}, vol.~37,
  pp. 246--248, 1960.

\bibitem{LiMD:19}
C.~Li, S.~F. Mao, and Z.~J. Ding, ``Time-dependent characteristics of secondary
  electron emission,'' \emph{J. Appl. Physics}, vol. 125, no. 024902, Jan.
  2019.

\bibitem{Joy2008}
D.~C. Joy, ``Noise and its effects on the low-voltage sem,'' in
  \emph{Biological Low-Voltage Scanning Electron Microscopy}, H.~Schatten and
  J.~B. Pawley, Eds.\hskip 1em plus 0.5em minus 0.4em\relax Springer New York,
  2008, pp. 129--144.

\bibitem{Joy1996}
D.~C. Joy, C.~S. Joy, and R.~D. Bunn, ``Measuring the performance of scanning
  electron microscope detectors,'' \emph{Scanning}, vol.~18, pp. 533--538,
  1996.

\bibitem{Joy2006}
D.~C. Joy, ``A database on electron-solid interactions,'' \emph{Scanning},
  vol.~17, pp. 270--275, Dec. 2006.

\bibitem{Uchikawa1992}
Y.~Uchikawa, K.~Gouhara, S.~Yamada, T.~Ito, T.~Kodama, and P.~Sardeshmukh,
  ``Comparative study of electron counting and conventional analogue detection
  of secondary electrons in sem,'' \emph{J. Electron Microscopy}, vol.~41, pp.
  253--260, 1992.

\bibitem{Frank2005}
L.~Frank, ``Noise in secondary electron emission: the low yield case,''
  \emph{J. Electron Microscopy}, vol.~54, no.~4, pp. 361--365, Aug. 2005.

\bibitem{Novak2009a}
L.~Novák and I.~Müllerová, ``Single electron response of the
  scintillator-light guide-photomultiplier detector,'' \emph{J. Microscopy},
  vol. 233, pp. 76--83, 2009.

\bibitem{TimischlDN:12}
F.~Timischl, M.~Date, and S.~Nemoto, ``A statistical model of signal---noise in
  scanning electron microscopy,'' \emph{Scanning}, vol.~34, no.~3, pp.
  137--144, 2012.

\bibitem{MartinK:62}
D.~C. Martin and S.~K. Katti, ``Approximations to the {N}eyman type {A}
  distribution for practical problems,'' \emph{Biometrika}, vol.~18, no.~3, pp.
  354--364, Sep. 1962.

\bibitem{croon2002line}
J.~Croon, G.~Storms, S.~Winkelmeier, I.~Pollentier, M.~Ercken, S.~Decoutere,
  W.~Sansen, and H.~Maes, ``Line edge roughness: characterization, modeling and
  impact on device behavior,'' in \emph{Digest. International Electron Devices
  Meeting,}, 2002, pp. 307--310.

\bibitem{MoulinV:19}
P.~Moulin and V.~V. Veeravalli, \emph{Statistical Inference for Engineers and
  Data Scientists}.\hskip 1em plus 0.5em minus 0.4em\relax Cambridge Univ.
  Press, 2019.

\bibitem{CorlessGHJK:96}
R.~M. Corless, G.~H. Gonnet, D.~E.~G. Hare, D.~J. Jeffrey, and D.~E. Knuth,
  ``On the {L}ambert {$W$} function,'' \emph{Adv. Comput. Math.}, vol.~5,
  no.~1, pp. 329--359, Dec. 1996.

\bibitem{Yamada1990}
S.~Yamada, T.~Ito, K.~Gouhara, and Y.~Uchikawa, ``Electron counting for
  secondary electron detection in {SEM},'' \emph{Scanning}, vol.~12, pp. I--28
  -- I--29, 1990.

\bibitem{Yamada1990a}
------, ``Secondary electron counting images in {SEM},'' in \emph{Proc. XIIth
  Int. Congress for Electron Microscopy}, 1990, pp. 402--403.

\bibitem{Yamada1991}
------, ``Electron-count imaging in {SEM},'' \emph{Scanning}, vol.~13, pp.
  165--171, 1991.

\bibitem{Yamada1991a}
------, ``High-speed electron counting system for {TV}-scan rate {SE} images of
  {SEM},'' in \emph{Proc. 49th Ann. Meeting of the Electron Microscopy Society
  of America}, 1991, pp. 512--513.

\bibitem{AGARWAL2021}
A.~Agarwal, J.~Simonaitis, and K.~K. Berggren, ``Image-histogram-based
  secondary electron counting to evaluate detective quantum efficiency in
  sem,'' \emph{Ultramicroscopy}, vol. 224, p. 113238, 2021.

\bibitem{Agarwal:2023-U}
A.~Agarwal, J.~Simonaitis, V.~K. Goyal, and K.~K. Berggren, ``Secondary
  electron count imaging in {SEM},'' \emph{Ultramicroscopy}, vol. 254, no.
  113662, 2023.

\bibitem{Seiler1983}
H.~Seiler, ``Secondary electron emission in the scanning electron microscope,''
  \emph{J. Appl. Phys.}, vol.~54, p.~R1, 1983.

\bibitem{Ishikawa2014}
R.~Ishikawa, A.~R. Lupini, S.~D. Findlay, and S.~J. Pennycook, ``Quantitative
  annular dark field electron microscopy using single electron signals,''
  \emph{Microsc. Microanal.}, vol.~20, pp. 99--110, Feb. 2014.

\bibitem{Sang2016}
X.~Sang and J.~M. Lebeau, ``Characterizing the response of a scintillator-based
  detector to single electrons,'' \emph{Ultramicroscopy}, vol. 161, pp. 3--9,
  2016.

\bibitem{McMullan2016}
G.~McMullan, A.~Faruqi, and R.~Henderson, ``Direct electron detectors,'' in
  \emph{The Resolution Revolution: Recent Advances In cryoEM}, ser. Methods in
  Enzymology, R.~Crowther, Ed.\hskip 1em plus 0.5em minus 0.4em\relax Academic
  Press, 2016, vol. 579, pp. 1--17.

\end{thebibliography}
    
\end{document}